\documentclass[aps,pre,reprint,superscriptaddress, nourl, noeprint, nodoi]{revtex4-2}

\usepackage[version=4]{mhchem}
\usepackage{graphicx}
\usepackage{xcolor}
\usepackage{dcolumn}
\usepackage{bm}
\usepackage[utf8]{inputenc}
\usepackage{upgreek}
\usepackage{appendix}
\usepackage{array}
\usepackage{booktabs}
\usepackage{makecell}
\usepackage[textsize=tiny]{todonotes}

\newcommand{\nobarfrac}{\genfrac{}{}{0pt}{}}

\begin{document}

\title{Polydisperse polymer fractionation between phases}

\author{J. Pedro de Souza}
\affiliation{Omenn-Darling Bioengineering Institute, Princeton University, Princeton, NJ, 08544}

\author{William M. Jacobs}
\email{wjacobs@princeton.edu}
\affiliation{Department of Chemistry, Princeton University, Princeton, NJ 08544, USA}

\author{Howard A. Stone}
\email{hastone@princeton.edu}
\affiliation{Department of Mechanical and Aerospace Engineering, Princeton University, Princeton, NJ, 08544}

\begin{abstract}
Polymer mixtures fractionate between phases depending on their molecular weight.  Consequently, by varying solvent conditions, a polydisperse polymer sample  can be separated between phases so as to achieve a particular molecular weight distribution in each phase. In principle, predictive physics-based theories can help guide separation  design and interpret experimental phase-diagram and fractionation measurements. Even so, applying the standard Flory-Huggins model can require numerical computations that hamper the predictions considering the full molecular weight distribution. Here, we apply a recently-derived exact analytical solution of multi-component Flory-Huggins theory for polydisperse homopolymers to understand the principles of polymer fractionation for common molecular weight distributions. Consistent with previous studies, the method highlights  the sensitivity of polymer fractionation to the shape, and in particular the tails, of this distribution. Our results provide a systematic approach to evaluate the full molecular weight distribution in phase coexistence calculations over the possible composition space.
\end{abstract}

\maketitle

\begin{figure*}[htbp!]
\centerline{\includegraphics[scale=0.65]{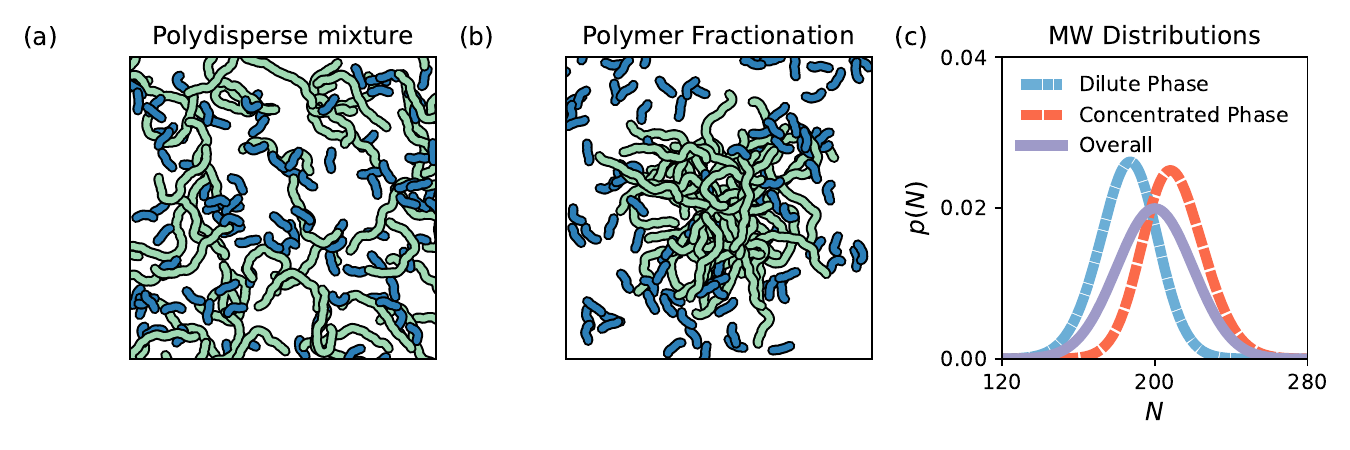}}
\caption{Polymer fractionation between phases. (a) A uniformly mixed polydisperse sample of long polymers (green) and short polymers (blue). (b) When this polydisperse mixture phase separates, longer polymers are preferentially concentrated in the concentrated phase and shorter polymers are preferentially accumulated in the dilute phase. (c) Generic fractionation behavior. The purple curve corresponds to the overall MW distribution of a ``dry" polymer sample exhibiting a normal distribution, the blue corresponds to the dilute B phase, and the red corresponds to the concentrated A phase when the polymer is placed in solvent. The MW distributions are expressed as a mass fraction probability density, $p(N_i)$, where $N_i$ is the number of monomers in component $i$, computed for a particular interaction strength and condensed volume fraction.   %{\color{red} panels need labels; couldn't panel c just be labeled generic since it is trying to just illustrate the idea.}
}
\label{Figure1}
\end{figure*}

\section{Introduction}

The fractionation of polymers refers to the re-distribution of polymers of different types {\color{black} and/or} lengths between phases~\cite{flory1953principles, cantow2013polymer, francuskiewicz2013polymer}. The simplest scenario is when all polymers are chemically identical in their monomer composition, but differ in their length or molecular weight (MW), which is called polydispersity~\cite{hiemenz2007polymer, gentekos2019controlling}. Virtually all polymer samples exhibit some degree of polydispersity arising from their synthesis conditions~\cite{odian2004principles, billmeyer1965characterization, whitfield2019tailoring}. 

If placed in a poor solvent, the polydisperse polymer will phase separate into a condensed (concentrated) polymer phase and a dilute phase composed primarily of solvent. The two phases will differ in their MW distributions~\cite{mencer1988efficiency}, since the entropy of demixing differs based on polymer length. As an example, a polydisperse polymer sample separates into a dilute phase, which is enriched in shorter polymers, and a concentrated phase, which is enriched in longer polymers, as illustrated in Figure \ref{Figure1}.

%TC:ignore

%TC:endignore

% \todo{Do you want to refer to a figure here?}

From synthetic to naturally occurring polymers, the MW polydispersity depends on the conditions of the polymerization reaction and the degree of homogeneity of a sample. For example,  ``living'' polymers, which are formed by monomer addition without termination, follow a modified Poisson distribution~\cite{flory1953principles, hiemenz2007polymer}. The Flory-Schulz distribution, which exhibits an extended tail, is frequently observed for linear polymers synthesized by step-growth reactions. Polymers synthesized by chain growth reactions have more diverse MW distributions \cite{odian2004principles}, which can be generically fit to normal or log-normal distributions~\cite{wang2023precise}. Naturally occurring biopolymers exist in complex polydisperse mixtures~\cite{van2023quantitative}.  The four common MW distributions that we study  are given in Table I.

Polydispersity strongly influences phase partitioning and composition in a wide range of systems, from engineered materials to biological mixtures. In industrial applications, a particular polymer MW distribution is needed to create materials of desired mechanical~\cite{ward2012mechanical}, rheological~\cite{bird1987dynamics},  thermal ~\cite{godovsky2012thermophysical} and thermoresponsive~\cite{ward2011thermoresponsive}, electrical~\cite{blythe2005electrical}, or optical ~\cite{higashihara2015recent} properties~\cite{gentekos2019controlling, whitfield2019tailoring}. Thus, by carefully choosing a solvent and/or processing temperature, polymer fractionation may be a simple route to concentrate particular polymer MWs into either a condensed or dilute phase. Similarly, naturally occurring biopolymers within cells may be fractionated between  condensed and dilute phases by forming biomolecular condensates~\cite{shin2017liquid, banani2017biomolecular, jacobs2023theory}. For example, certain RNA granules exhibit a higher propensity to concentrate longer RNA with more exposed sequences~\cite{tian2020rna}.

% \todo{I think that you can compress the previous three paragraphs and make the synthetic vs bio comments more parallel to one another.}

\begin{table}[b!]
\begin{ruledtabular}
\setlength{\tabcolsep}{0pt}
\renewcommand{\arraystretch}{1}
\resizebox{\columnwidth}{!}{%
\begin{tabular}{c c c}
& & \\
MW Distribution & $p(N_i)$ & Weight Average $\langle N \rangle$\\ & & \\\hline
& & \\
Normal Distribution &
\makecell[c]{$\frac{\exp\left(-\frac{(N_i-\langle N \rangle)^2}{2N_\sigma^2}\right)}{\sqrt{2\pi N_\sigma^2}}$} &
\makecell[c]{$\langle N \rangle$} \\
& & \\
Modified Poisson &
\makecell[c]{$\frac{\lambda}{\lambda+1} \frac{N_i e^{-\lambda} \lambda^{N_i-2}}{(N_i-1)!}$} &
\makecell[c]{$\lambda+2-\frac{1}{\lambda+1}$} \\
& & \\
Bimodal &
\makecell[c]{$\nobarfrac{\eta \, p_\mathrm{norm}(N_i, N_1, N_\sigma)}{+(1-\eta) p_\mathrm{norm}(N_i, N_2, N_\sigma)}$} &
\makecell[c]{$\eta\, N_1+(1-\eta)N_2$} \\
& & \\
Flory-Schulz &
\makecell[c]{$N_i(1-b)^2 b^{N_i-1}$} &
\makecell[c]{$\frac{1+b}{1-b}$} \\
& & \\
\end{tabular}%
}
\end{ruledtabular}
\caption{\label{tab:example}The MW distributions studied in this work, expressed as a mass fraction of polymer chains of length $N_i$. The value of $p(N_i)$ is the known probability density of component $i$ in the ``dry'' polymer sample without solvent. For the normal distribution, $\langle N\rangle$ describes the center of the distribution, while $N_\sigma$ describes the standard deviation. In the modified-Poisson distribution, the $\lambda$ parameter encapsulates both the average and variance of the distribution. For the bimodal distribution, the parameters $N_1$ and $N_2$ characterize the centers of the peaks in the MW distribution, while $N_\sigma$ describes the standard deviation, and $\eta$ describes the fractional amount of the peak centered at $N_1$. The parameter $b$ describes the extent of reaction $0<b<1$ for Flory-Schulz.}
\end{table}

Theoretical model predictions of polymer fractionation explain phase behavior at a quantitative level and aid in process design. Even so, phase coexistence calculations can be computationally demanding for multi-component mixtures~\cite{heidemann1995instability, jacobs2023theory, van2023predicting}. While simplifications for evaluating fractionation efficiency of homopolymers can be made~\cite{flory1953principles}, phase coexistence calculations still require iterative numerical solutions that consider all species, even those occurring in minute quantities in the tails of the MW distribution. 

Given the importance of polydispersity in polymer solutions~\cite{sollich2001predicting}, the numerical calculations of polydisperse phase diagrams go back more than half a century to early work of Koningsveld and coauthors~\cite{koningsveld1969phase}. There, they performed computations predicting the fractionation from different distributions under a variety of conditions, usually in the interest of optimizing a criterion for a given fractionation efficiency~\cite{koningsveld1970preparative, koningsveld2001polymer}, and occasionally including the composition dependence of the Flory interaction parameter. Early on, \v{S}olc recognized that the form of the tail of the distribution was an essential determinant in the behavior of coexisting cloud and shadow curves~\cite{solec1970cloud, vsolc1975cloud}, the composition curves that demarcate the onset of phase separation on the dilute and condensed branch of the phase diagram. If a distribution's tail decays more slowly than exponentially with MW, then above a critical interaction strength, the dilute branch of the cloud curve goes towards zero concentration, and the corresponding shadow curve adopts a universal form corresponding to the limit of vanishing osmotic pressure. These analyses applied primarily to the cloud and shadow curves of the polydisperse mixture. Furthermore, three-phase coexistence has been predicted~\cite{tompa1949phase, koningsveld2001polymer} and found experimentally~\cite{koningsveld1967liquid} for samples containing disparate lengths of polymers, albeit in narrow ranges of interaction strengths (or temperature) near the critical point.

The general problem of phase coexistence in polydisperse mixtures is complicated, and existing methods, while well developed, are not easily applied to arbitrary polydispersity distributions. Among these, moment free energy approaches~\cite{sollich1998projected, sollich2001predicting, fasolo2003equilibrium, fasolo2004fractionation, speranza2003isotropic2, wilding2005finite, patyukova2021phase} offer an elegant and versatile framework for treating polymer length polydispersity and more general forms of heterogeneity, including random copolymer composition and related features. In practice, however, their application still generally relies on iterative numerical solution procedures. Despite the comprehensive body of work on polydisperse fractionation and phase coexistence, in practical applications, the effects of polydispersity are often neglected, presumably due to either a lack of information on the MW distribution or the need for solving systems of nonlinear equations numerically over the composition space. These limitations underscore the need for approaches that make quantitative treatment of polydispersity more practical in routine applications.

We address this challenge by applying a new calculation technique that vastly simplifies two-phase coexistence calculations. In our recent work~\cite{deSouza2024exact}, we found an exact implicit analytical solution to two-phase coexistence within the Flory-Huggins model describing a polydisperse polymer mixture with any number of components. The exact solution allows us to evaluate the interplay of polydispersity and phase separation at high numerical accuracy. Further, we can self-consistently predict the two-phase fractionation of polymers over the full compositional range with minimal computational effort.  

% : “when is polydispersity important to consider for phase separation of polymers with common MW distributions?”  “how does polydispersity of various types affect phase coexistence?” and  “How are polymers fractionated at different conditions? 

% {\color{red} should we remark that we have assumed only two phase in our original work?}{\color{blue} For a polydisperse polymer in \textbf{length} only, it is only possible to get two phase coexistence, as far as I know. This is because the partitioning of all components other than solvent has the exact same sign. This is also assumed in Flory's textbook chapter section on ``polymer fractionation," and I think it is generally true.}

% Evaluating the exact solution requires sums over all molecular components, requiring information from the entire MW distribution. While these sums are readily available for numerical evaluation, it may be advantageous to generate simpler formulas based on a smaller set of descriptive characteristics of the distribution.  For example, one may desire to predict the fractionation of polymers based only on the average and standard deviation of the MW distribution. More generally, we may wish to estimate phase coexistence by using the moments of the distribution, $m_\ell$,
% that are tabulated or calculable for common distributions.

Here, we explore the effects of polydispersity for common MW distributions, employing our analytical method within the Flory-Huggins theory that simply requires evaluating a set of analytical formulas. We first provide background on the construction of the Flory-Huggins theory and the analytical method employed to calculate two-phase coexistence. Next, we apply the method to common MW distributions, showing how the MW distribution shifts between phases, how the phase diagram itself shifts due to polydispersity, and how this varies with the total polymer composition. Consistent with previous predictions~\cite{sollich2001predicting, koningsveld2001polymer}, we find that the extent of polydispersity, along with the tails of the distribution, can strongly influence the phase diagram and resulting fractionation between phases. Within the analytical method, we demonstrate how slowly decaying tails in the MW distribution lead to a strong propensity towards phase separation dominated by the longest polymers present, highlighting that for such common MW distributions, phase coexistence is relevant at even minute concentrations. Also, we revisit the appearance of three-phase coexistence, showing the crossover signatures from the cloud curve that indicate these regimes.

\section{Theoretical model}
Within the framework of Flory-Huggins theory, we assume that the enthalpic interactions between monomers are the same regardless of polymer chain length. The free energy density, $f$, of polydisperse polymers of length $N_i$, filling a lattice with volume $v$ per monomer or solvent molecule, is
\begin{equation}
  \frac{fv}{k_BT}=\sum_{i=1}^{M}\frac{1}{N_i}\phi_i\ln(\phi_i)+(1-\phi_t)\ln(1-\phi_t)-\chi\phi_t^2,
\end{equation}
where $\phi_i$ is the volume fraction of component $i$, $\phi_t=\sum_{i=1}^{M}\phi_i$ is the total polymer volume fraction,  and $\chi$ represents the effective interactions of the polymer segments with the solvent.  The summation over $i$ accounts for all $M$ polymeric components of differing lengths. 
% \todo{Perhaps note the two-phase assumption here.}

% {\color{blue} Note that since we are assuming constant volume and working with the Helmholtz free energy, the $\chi$ term is actually formally an internal energy, not an enthalpy. It's a bit nuanced, but I think that this is more consistent with the lattice-based derivation of Flory-Huggins. Technically speaking, the term may also have effective contributions from entropy changes in the solvent orientations or hydration of polymers.}

% \todo{`Weight average' may be unclear; define in caption?.  Use a symbol for WA, and then refer to it throughout instead of $N_1$, etc.}

% {$\begin{aligned}
%         & f p_\mathrm{norm}(N_i, N_1, N_\sigma) \\
%         & +(1-f) p_\mathrm{norm}(N_i, N_2, N_\sigma)
%         \end{aligned}$} & 
%         \makecell[c]{$(N_i(1-b)^2 b^{N_i-1})$}

Here, we will start by assuming that phase separation results in coexistence between two phases. 
 At thermodynamic equilibrium, each component $i$ must have equal chemical potential, $\mu_i$, in each phase, and the osmotic pressure, $\Pi$, between phases $A$ (condensed) and $B$ (dilute) must be identical,
\begin{subequations}
    \begin{equation}\label{eq:muEq}
        \mu_i(\Phi_A)=\mu_i(\Phi_B)
    \end{equation}
    \begin{equation}\label{eq:PiEq}
        \Pi(\Phi_A)=\Pi(\Phi_B),
    \end{equation}
\end{subequations}
where $\Phi$ is a vector containing the volume fractions of each component $i$. The chemical potential is derived as
\begin{equation}
    \frac{\mu_i}{k_B T}=\frac{1}{N_i}\ln(\phi_i)-\ln(1-\phi_t)-2\chi \phi_t,
    \label{ChemPotentialEqn}
\end{equation}
and the osmotic pressure is
\begin{equation}
    \frac{\Pi v}{k_B T}=-\ln(1-\phi_t)-\chi\phi_t^2 -\phi_t+\sum_{i=1}^{M}\frac{\phi_i}{N_i}.
\end{equation}

The mass balances between the phases for each component $i$ provide an additional constraint, 
\begin{equation}\label{eq:MassBalEq}
    \bar{\phi}_i=\nu \phi_{iA}+(1-\nu)\phi_{iB},
\end{equation}
with $\nu$ the volume fraction of the condensed phase and $\bar{\phi}_i$ the total polymer $i$ volume fraction. 

% The system of equations (\ref{eq:muEq}), (\ref{eq:PiEq}), and (\ref{eq:MassBalEq}), along with knowledge of the overall molecular weight distribution, we have a nonlinear system of equations that scales with the number of components. However, this system is difficult to solve directly. 

With the system of equations (\ref{eq:muEq}), (\ref{eq:PiEq}), and (\ref{eq:MassBalEq}), along with knowledge of the overall molecular weight distribution, $p(N_i)$,
\begin{equation}\label{eq:DistributionEq}
    \frac{\bar{\phi}_i}{\bar{\phi_1}}=\frac{p(N_i)}{p(N_1)},
\end{equation}
we have a nonlinear system of $3M$ equations and $3M+2$ unknowns ($\phi_{iA}, \phi_{iB}, \bar{\phi}_i, \chi, \nu$) that scales with the number of components. The problem can be fully specified  by fixing two degrees of freedom, such as $\chi$ and $\bar{\phi}_1$. However, this system is difficult to solve directly. 

\subsection{Typical fractionation calculation approach}

% Flory ~\cite{flory1953principles} and Huggins ~\cite{huggins1967theoretical} each noted that the chemical equilibrium, by equating $\mu_i-\Pi$ (in our notation) for each specie, directly gives a relationship describing the fractionation of each polymer up to an undetermined parameter which was fitted to experimental data at each condition (SI Section S1). Yet, the approach does not provide an exact solution for phase coexistence,  and Flory even stated that ``A complete calculation of the phase equilibrium from the equations would be a staggering task''~\cite{flory1953principles}. Therefore,  the approach may be useful for data interpretation, but is not directly predictive or transferable between different experimental conditions. 

Building from earlier work of Flory~\cite{flory1953principles} and Huggins~\cite{huggins1967theoretical}, each noted that the chemical equilibrium directly gives a relationship describing the fractionation of each polymer up to an undetermined parameter, which was fitted to experimental data at each condition. By equating $\mu_i-\Pi$ (in our notation) for each specie, there is a direct relationship describing the fractionation of each polymeric component,
\begin{equation}\label{eq:OmegaEq}
    \frac{\phi_{iA}}{\phi_{iB}}=\exp(N_i\Omega ),
\end{equation}
with the function $\Omega$ defined as
\begin{equation}
    \Omega=(-1+2\chi)(\phi_{At}-\phi_{Bt})-\chi(\phi_{At}^2-\phi_{Bt}^2)+\sum_{i=1}^{M}\frac{\phi_{iA}-\phi_{iB}}{N_i}.
\end{equation}

Combined with a mass balance of polymer species $i$, the above treatment gives an exact formula for the fraction of polymer $i$ in the concentrated phase, $\theta_i$:
\begin{equation}
    \theta_i=\frac{\nu \phi_{iA}}{\nu\phi_{iA}+(1-\nu)\phi_{iB}}{}=\frac{\nu e^{N_i\Omega}}{1-\nu+\nu e^{N_i\Omega}}
\end{equation}
where, again, $\nu$ is the volume fraction of the condensed phase.

In applying the above fractionation formula, both Flory and Huggins treated $\Omega$ as an effective fitted constant and did not attempt to solve the fractionation or phase coexistence in its entirety. In fact, Flory stated that ``A complete calculation of the phase equilibrium from the equations would be a staggering task''~\cite{flory1953principles}.

% \todo{One might expect that you also need to specify the total polymer mass, but I think that this is a dependent variable in this formulation.  Still, it might be confusing because $\nu$ is not the (directly) experimentally controllable variable.}

While this mathematical treatment of fractionation at fixed $\Omega$ gives helpful information on the separation of polymers between phases, it does not answer the question of how the polydispersity of polymers themselves alter the phase diagram in terms of $\chi$. The value of $\Omega$, which is in principle a function of the concentrations of all species in both phases and the volume of the condensed phase, must be fitted to an experimental data set at each particular condition and solved for numerically within phase coexistence calculations~\cite{koningsveld1969phase}. This means that the above expression can be used to interpret fractionation data, but is not systematically transferable between different experimental conditions. 

In general, the polydisperse problem requires more careful analysis to describe the phase equilibrium over the full compositional range. Yet, the assumption of two-phase coexistence can allow for considerable simplification by eliminating variables down to two equations in two unknowns, starting first from the two-phase equilibrium calculation of $3M+2$ variables and $3M$ equations. Equation (\ref{eq:OmegaEq}) explicitly eliminates $M$ variables ($\phi_{iA}$), while adding one new variable $\Omega$ and one equation defining $\Omega$.  Equation (\ref{eq:MassBalEq}) in turn can be used to explicitly eliminate $M$ more variables ($\phi_{iB}$) and equation (\ref{eq:DistributionEq}) eliminates $M-1$ variables $\bar{\phi}_{i>1}$. So after the eliminations of variables, we are left with four unknowns ($\Omega$, $\bar{\phi}_1$, $\chi$, and $\nu$) and two equations (\ref{eq:OmegaEq} and \ref{eq:PiEq}). By specifying two variables, for example, $\nu$ and $\chi$ over their allowable range, our task becomes to solve for $\Omega$ and $\bar{\phi}_1$ numerically. While this procedure can be implemented using iterative methods going back to early works~\cite{koningsveld1968liquid_calculation}, we simplify the inversion of the problem directly using a variable transformation method, as follows.

% In general, the problem requires more careful numerical analysis to describe the phase equilibrium over the full compositional range. The problem for two phase coexistence allows for a simplification down to a system of two nonlinear equations in two unknowns. Equation \ref{eq:OmegaEq} explicitly eliminates $M$ variables ($\phi_{iA}$), while adding one new variable $\Omega$ and one equation defining $\Omega$.  Equation \ref{eq:MassBalEq} in turn can be used to explicitly eliminate $M$ more variables ($\phi_{iB}$).Equation \ref{eq:DistributionEq} eliminates $M-1$ variables $\bar{\phi}_{i>1}$. So after the eliminations of variables, we are left with four unknowns ($\Omega$, $\bar{\phi}_1$, $\chi$, and $\nu$) and two equations (\ref{eq:OmegaEq} and \ref{eq:PiEq}). By specifying two variables, for example, $\nu$ and $\chi$ over their allowable range, our task becomes to solve for $\Omega$ and $\bar{\phi}_1$.

\subsection{Mathematical Methods}
In this work, we employ an implicit substitution method to solve the two-phase coexistence problem exactly. The key idea is that we treat the value of $\chi$ implicitly, inverting the problem into one transcendental equation in one unknown composite variable, and then transform back to real compositions.  We highlight that Flory's remark about the exact solution rings true--- the method is algebraically involved, requiring mappings between composition coordinates, but nevertheless it is computationally efficient to generate the thermodynamic coexistence curves by direct evaluation of analytical formulas. The exact solution is also advantageous since the computations can be done with high numerical accuracy, a feature that may need special attention to implement with other approaches. We will demonstrate that this accuracy is important since subtle features of MW distributions can have a strong influence on the phase diagram.

To solve the thermodynamic calculation exactly for a polydisperse sample with the allowable degrees of freedom, we recently proposed an implicit substitution method that makes use of composite composition variables~\cite{deSouza2024exact}. We include details of the fractionation calculation below. Since the focus here is to apply the method to a variety of molecular weight distributions, we refer to ~\cite{deSouza2024exact} for details of the derivation.

% \todo{I'm not sure that the meaning of the equations/variables is clear.} 
Here, we will specify the component of index 1 to be the component closest to the mass average molecular weight of the overall polymer sample. To start, we frame the equations in terms of a differential partitioning variable, $y_i$, between the two phases, defined as
\begin{equation}
    y_i=\frac{\phi_{iA}-\phi_{iB}}{\phi_{iA}+\phi_{iB}}.
\end{equation}
$y_i$ is a measure of the partitioning of component $i$ and can also indicate the distance from the critical point. 
By setting $\mu_i-\mu_1$ equal to each other in each phase, we find
\begin{equation}\label{eq:eqyN}
    y_i=\tanh\left(\frac{N_i}{N_1}\tanh^{-1}(y_1)\right).
\end{equation}

Next, by solving each chemical potential equilibrium equation for $\chi$, and then subtracting a linear combination of them from the value of $\chi$ from the osmotic pressure equation, we can solve for all the volume fractions exactly. In order to do so, we reframe the equations in terms of the solvent partitioning, $z$,
\begin{equation}
    z=\frac{\phi_{tA}-\phi_{tB}}{2-\phi_{tA}-\phi_{tB}}
\end{equation}
and the set of relative partitioning variables, $w_i$,
\begin{equation}
    w_i=\frac{\phi_{iA}-\phi_{iB}}{\phi_{1A}-\phi_{1B}},
\end{equation}
comparing the partitioning of component $i$ to component 1. Here, we will independently set two variables:  (i) the partitioning of component 1, $y_1$, and (ii) the volume fraction of the condensed phase, $\nu$. These are two independent degrees of freedom that will specify the entire two-phase diagram.

\begin{figure*}[t!]
\centerline{\includegraphics[scale=0.55]{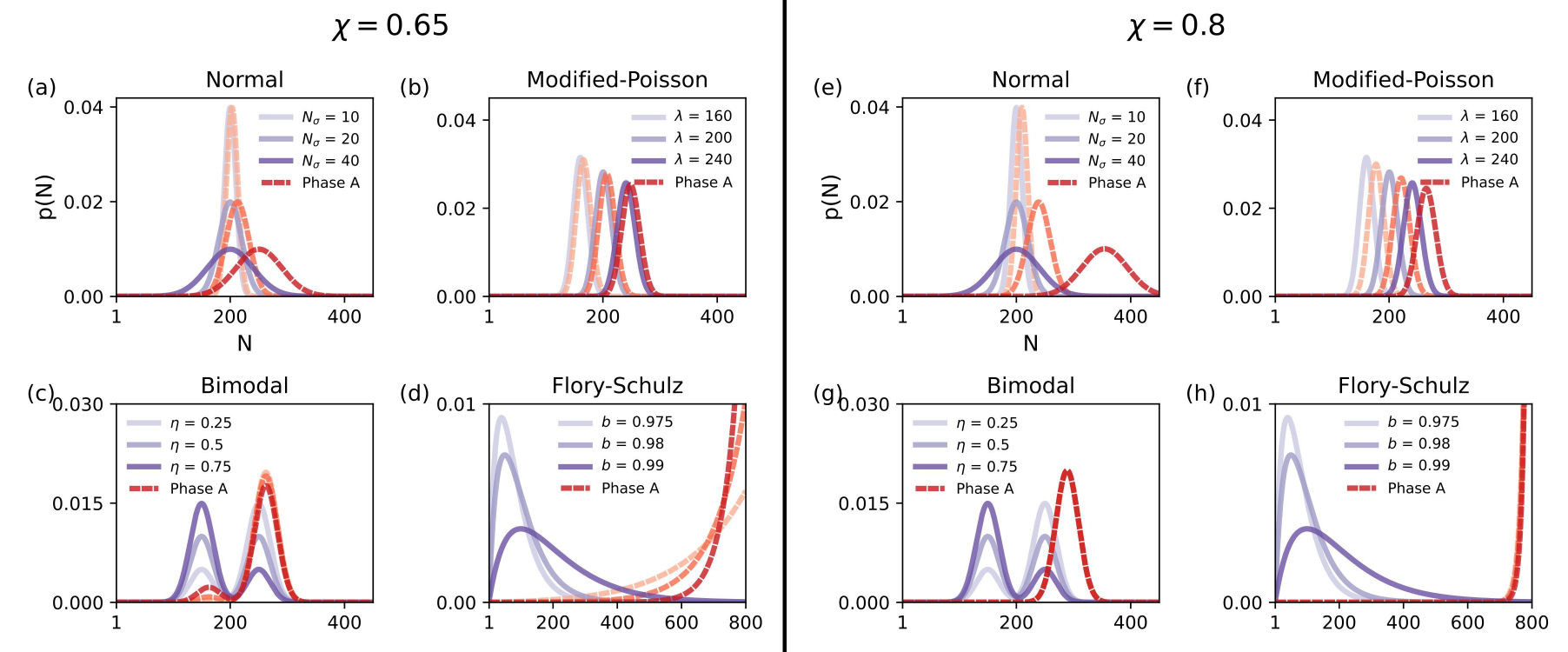}}
\caption{Fractionation at the cloud point ($\nu=0$). The MW distributions with (a-d) $\chi=0.65$  and 
(e-h) $\chi=0.8$  at the cloud point, when the condensed phase (A) first appears. The shade of the line designates a correspondence between the overall distribution (purple) and the condensed distribution (red). The distributions tested are (a,e) normal distribution centered at $N_1=200$ with varying standard deviation, $N_\sigma$,  (b,f) modified-Poisson distributions with varying $\lambda$ parameters, (c,g) bimodal distributions with $N_2=250$ and $N_1=150$ with $N_\sigma=20$, varying the fraction $\eta$ of the first peak, and (d,h) Flory-Schulz for varying degrees of polymerization, $b$. More polydisperse samples exhibit greater fractionation, and the fractionation is dominated by the tails of the distribution, with greater fractionation also at higher $\chi$.  }
\label{Figure2}
\end{figure*} 

From the mass balance for each component, and the mass balance of component 1, we can define $w_i$ in terms of $\nu$ and $y_1$,
\begin{equation}\label{eq:eqwN}
    w_i=\frac{p(N_i)}{p(N_1)}\frac{\left(\nu+\frac{1}{2 y_1}-\frac{1}{2}\right)}{\left(\nu+\frac{1}{2 y_i}-\frac{1}{2}\right)},
\end{equation}
where $p(N_i)$ is the known probability density of component $i$ in the ``dry'' polymer sample without solvent.

We can now account for the number of equations and unknown composite variables we must solve for. There are $M-1$ equations describing the set $w_i$ in equation (\ref{eq:eqwN}). The equilibrium constraints on each component's chemical potential and osmotic pressure give $M+1$ equations, for a total of $2M$ equations. In terms of the composite variables, the unknowns are the $M$ values of $y_i$, the $M-1$ values of $w_i$ (since $w_1=1$ automatically), as well as $z$, $\nu$, and $\chi$, for a total of $2M+2$ unknowns. Therefore, we have two degrees of freedom, which we use to independently specify $y_1$ and $\nu$ over their allowable range, then find all other composite variables.

By setting $y_1$ and $\nu$ and having knowledge of the overall molecular weight distribution $p(N_i)$, we can find the solvent partitioning $z$ exactly by enforcing osmotic equilibrium:
\begin{equation}
    z=h^{-1}\left(1+\frac{\sum_{i=1}^{M} w_i(h(y_i)-1)/N_i}{\sum_{i=1}^{M} w_i}\right),
\end{equation}where we have defined the  Flory-Huggins (FH) function $h(x)$ as
\begin{equation}
    h(x)=\frac{\tanh^{-1}(x)}{x}.
    \label{deSouzaJacobsStone-FloryHugginsEqn}
\end{equation}
Its inverse, $h^{-1}()$, can be efficiently computed by lookup tables or by a series representation, which is derived in Appendix B.

Finally, equipped with the value of $z$, we can map the composite composition coordinates back to real composition space,
% \todo{It is unclear how you solve for $\chi$, since it's not in the preceding equations. Update---I answered my own question (see below), but I think this section needs to be reworked.}
\begin{equation}
    \{y_i, z, w_i, \nu\}\rightarrow \{\phi_{iA}, \phi_{iB}, \chi\}.
\end{equation}
Explicitly, the mappings take the form
\begin{subequations}
    \begin{equation}
        \phi_{iA}= \frac{1}{2}\beta_i(1+y_i)
    \end{equation}
    \begin{equation}
        \phi_{iB}=\frac{1}{2}\beta_i(1-y_i)
    \end{equation}
\end{subequations}
and
\begin{equation}
    \chi=\frac{2\tanh^{-1}(z)+\sum_{i=1}^{M}\left(\frac{1}{N_i}-1\right)\beta_i y_i}{\gamma_0},
\end{equation}
%For the polydisperse sample, 
where the constants $\gamma_0$ and $\beta_i$ are defined as
\begin{equation}
    \gamma_0=\sum_{i=1}^{M} \sum_{j=1}^{M}\left(\phi_{iA}\phi_{jA}-\phi_{iB}\phi_{jB}\right)
\end{equation}
and
\begin{equation}\label{eq:beta_defined}
    \beta_i=\frac{2 z w_i}{y_i\sum_{j=1}^{M}\left[w_j\left(1+\frac{z}{y_j}\right)\right]}.
\end{equation}

For the polydisperse system under study, one may notice that all the real composition variables can be computed based on three weighted sums of the molecular weight distribution for a fixed $y_1$ and $\nu$. These weighted sums can be expressed as $I_k$ ($k=1,2,3$):
        \begin{equation}
        I_k=\sum_{i=1}^{i=M}\frac{p(N_i)}{p(N_1)} Q_k(N_i),
        \end{equation}
where the weighting functions $Q_k(N_i)$ are defined as
    \begin{subequations}
        \begin{equation}
        Q_1(N_i)=\frac{\nu+\frac{1}{2 y_1}-\frac{1}{2}}{\nu+\frac{1}{2 y(N_i)}-\frac{1}{2}}
        \end{equation}
        \begin{equation}
            Q_2(N_i)=Q_1(N_i)\frac{h(y(N_i))-1}{N_i}
        \end{equation}
        \begin{equation}
        Q_3(N_i)=\frac{Q_1(N_i)}{y(N_i)}.
        \end{equation}
    \end{subequations}

    Recasting the computed variables in terms of these weighted sums, the volume fraction in the condensed phase (A) for each species is
\begin{equation}\label{eq:eqphiA}
\phi_{iA}=\phi_A(N_i)=\frac{zp(N_i)Q_1(N_i)\left(1+y(N_i)\right)}{p(N_1)y(N_i)\left(I_1+zI_3\right)},
\end{equation}
and in the dilute phase (B) the volume fraction is
\begin{equation}\label{eq:eqphiB}
\phi_{iB}=\phi_B(N_i)=\frac{zp(N_i)Q_1(N_i)\left(1-y(N_i)\right)}{p(N_1)y(N_i)\left(I_1+zI_3\right)}.
\end{equation}
The partitioning of the solvent, $z$, can be written in the form
\begin{equation}
    z=h^{-1}\left(1+\frac{I_2}{I_1}\right), 
\end{equation}where $h(x)$ is defined in equation (\ref{deSouzaJacobsStone-FloryHugginsEqn}).
The total polymer volume fractions in each phase are then
\begin{subequations}
    \begin{equation}
        \phi_{tA}= \frac{z(I_3+I_1)}{I_1+zI_3}
    \end{equation}
    \begin{equation}
        \phi_{tB}=\frac{z(I_3-I_1)}{I_1+zI_3},
    \end{equation}
\end{subequations}
and the total polymer volume fraction, $\bar{\phi}$ in the solution, can be computed as
\begin{equation}\label{eq:eqbarphidef}
    \bar\phi=\nu \phi_{tA}+(1-\nu)\phi_{tB}.
\end{equation}
The physical variable that is often controlled and easily measured is the total volume fraction of polymer that is added to the solution, $\bar\phi$. Therefore, while we compute the phase diagram by enumerating values of $\nu$ between 0 and 1, we map this to $\bar{\phi}$ using (\ref{eq:eqbarphidef}) for more physical results.  

Taken together, the system of equations outlined above can be evaluated directly in closed form. We provide python codes ~\cite{GitCode} to perform calculations including accurate evaluations of special functions that can be applied to arbitrary MW distributions. As we will demonstrate, the method easily scales for evaluating full phase diagrams that would otherwise require iterative solution methods at each discrete point over two degrees of freedom.

\begin{figure}[b!]
\centerline{\includegraphics[scale=0.45]{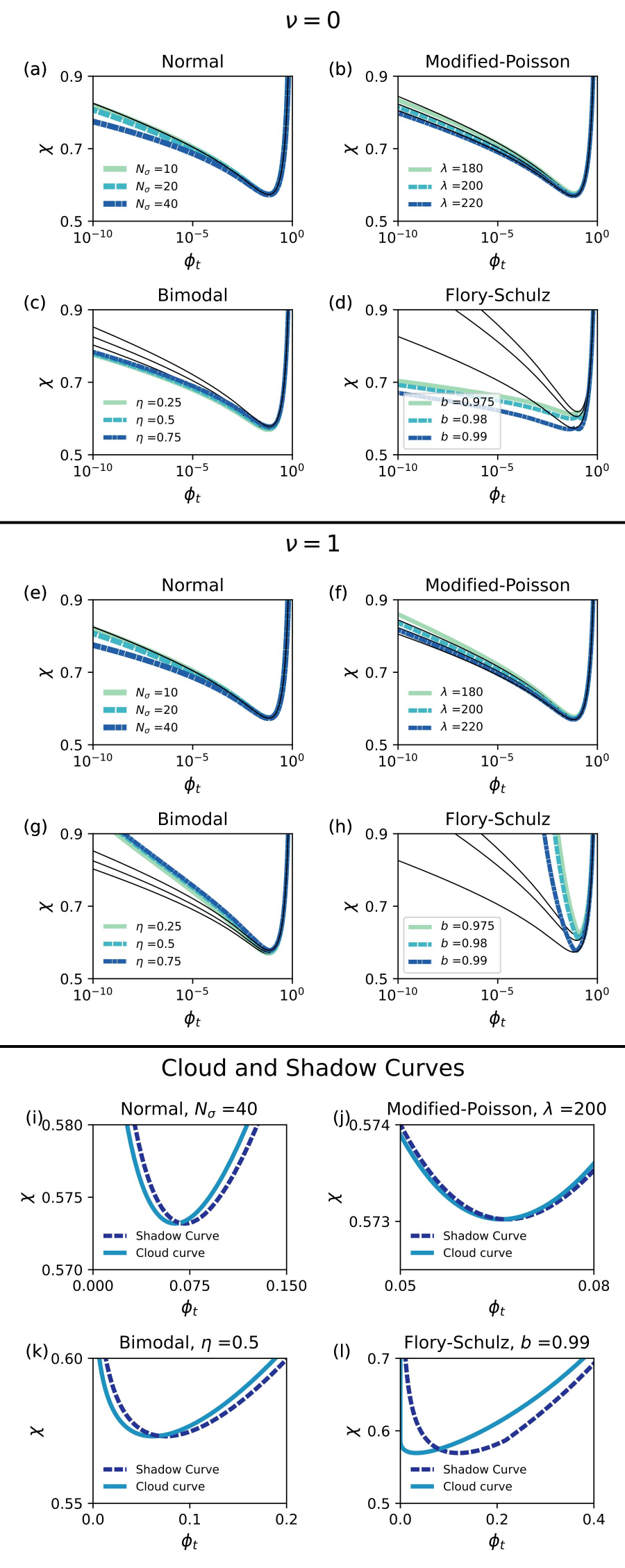}}
\caption{Polydispersity shifts the phase boundary. 
The two-phase coexistence curves are plotted as $\chi(\phi_t)$, evaluated at the cloud point (a-d) where an infinitesimal amount of condensed phase is formed ($\nu=0$), or at (e-h) where an infinitesimal amount of the dilute phase is formed ($\nu=1$). The cloud curve and shadow curve are replotted (i-e) with rescaled x-axis for one distribution in each panel. Due to the polydispersity, the cloud and shadow curves do not overlap.  The overall (dry) MW distributions are the same as in Fig.~\ref{Figure2}, with (a,e,i) a normally distributed polymer sample with varying standard deviation, (b,f,j) a modified-Poisson distributed sample, (c,g,k) a bimodal distributed polymer sample, and (d,h,l) a Flory-Schulz distributed sample with varying degrees of polymerization. \color{black}The thin black lines correspond to the behavior of a monodisperse solution of weight-average molecular weight.\color{black}  }
\label{Figure3}
\end{figure}

\section{Results and Discussion}

% \todo{Could these numerically exact results be computed without your method?}

% \todo{Do you want to say something like ``The rest of the paper is organized as follows.''? Based on the first sentence, I thought you were going to jump into cloud pt calculations, but then you gave an outline.}
The results are organized as follows. Using our analytical method, we first evaluate the fractionation of different distributions near the cloud point---defined as the point when an infinitesimal amount of condensed phase forms. We then examine how the molecular weight distribution can influence the observed phase diagram, which is typically measured at the cloud point. The complete phase diagrams are calculated for polydisperse samples over the full polymer/solvent composition range to identify how the phase coexistence shifts as the condensed phase volume changes. Finally, using this method we demonstrate the appearance of three-phase coexistence for samples with strong disparities in polymer length.

\subsection{Fractionation analysis}
First, we calculate the MW distribution in the condensed phase at the cloud point ($\nu=0$) with $\chi=0.65$ (a-d) and $\chi=0.8$ (e-h) for the four different  distributions listed in Table I, as shown in Fig.~\ref{Figure2}. At the cloud point, the overall distribution of the polymer sample (purple) matches the distribution in the dilute phase, but differs from the condensed phase (red). The parameterizations of the overall distributions are chosen such that the weight average chain length, $\langle N\rangle$, is $\approx 200$ for at least one of the curves from each distribution ($\langle N\rangle=200$ in (a,e), $\lambda=200$ in (b,f), $\eta=0.5$ in (c,g) and $b=0.99$ in (d,h)). These curves at constant $\langle N\rangle$ can be compared directly  to examine the effect of the MW distribution. 

%TC:ignore

%TC:endignore

% \todo{Fig 2: Fix panel labels in caption. Also, latex is telling me that Figure2 is defined twice; check to make sure your crossrefs are correct if you are using them.}

% \begin{figure*}[htbp!]
% \centerline{\includegraphics[scale=0.6]{polymer_frac_figs/Figure_3B.png}}
% \caption{Full phase diagrams for different MW distributions. (a-c) Results for a normally distributed sample with $N_1=200$ and $N_\sigma=20$. (a) The overall phase diagram surface $\chi(\phi_t, \bar{\phi})$, where $\phi_t$ is the sum of all species in a given phase and $\bar{\phi}$ is the overall volume fraction of polymer added to the solution. The values are saturated at the bounds of the color bar. (b-c) The MW distribution in the dilute and condensed phases, $p(N)$, shifts as polymer is added to solution, changing $\bar{\phi}$. (d-f) The analogous plots as (a) for different MW distributions. These include (d) a Flory-Schulz sample with $b=0.98$, (e) a Modified Poisson sample with $\lambda=200$, and (f) a bimodal distribution with $\eta=0.5$. {\color{red} panels need labels}}
% \label{Figure3}
% \end{figure*} \todo{I think that the full phase diagrams could be more easily understood if they were paired with the panels in Fig 2 (with corresponding points on the binodal indicated).  ** Could you indicate the line of cloud points on this plot?  ** See my comment above about panels b and c.}

Next, we examine the fractionation predictions for $\chi=0.65$. In Fig.~ \ref{Figure2}(a), by comparing normally-distributed samples of different standard deviation, $N_\sigma$, but with the same mean, $\langle N\rangle$, we demonstrate that greater polydispersity leads to greater fractionation from the high MW tails of the distribution. Therefore, the condensed phase from a sample with greater variance will have a greater mean MW. In contrast, the modified Poisson distribution in Fig.~ \ref{Figure2}(b) exhibits sharper, more uniform peaks than the other distributions~\cite{Note1}. There is relatively small translation in the MW distribution in the condensed phase relative to the overall distribution, although this translation increases with increasing value of $\lambda$.

The role of the highest peak in the overall MW distribution is clearly apparent for the bimodal distribution shown in Fig.~\ref{Figure2}(c). No matter the ratio of the two peaks in the bimodal distribution, the MW distribution in the condensed phase is dominated by the higher MW species and centered on the right arm of the higher MW peak.  
% \todo{Can this case be related to the long tail argument for the FS distribution?}

 The Flory-Schulz distribution, shown in Fig.~\ref{Figure2}(d), is unique from the other distributions studied here in that the tail decays slower than exponentially. As a result, the  fractionation at the cloud point is dominated by the longest polymer included in the sample, as we prove mathematically in the Appendix A. For our sample calculations, we truncate the distributions at $N=800$, and the distributions are therefore peaked at that value. If longer polymers are included, then the condensed MW distribution moves nearer to the maximum present chain length  at the cloud point. These results emphasize that the shape and decay of the molecular distribution  determine the fractionation behavior, particularly near the cloud point.

 % \todo{I suggest reordering the figure panels and discussion so that you cover well behaved examples first and then move to the strong fractionation cases.  If you do this, then I would emphasize the exponential tail versus long tail part at the beginning of this passage.}

The results demonstrated in Fig.~\ref{Figure2}(a-d) are extracted at the cloud point at one value of $\chi=0.65$. In Fig.~\ref{Figure2}(e-h), we demonstrate that fractionation is stronger as $\chi$ increases since there is a stronger driving force for demixing, consistent across all distributions tested. Furthermore, as polymer is added to solution beyond the cloud point, eventually the dry and condensed-phase distributions become similar, as fractionation of low MW components into the dilute phase eventually dominates.

% \todo{Do the effects strengthen or diminish as you move away from the cloud point?}

% Further, the shape of the MW distributions depend sensitively on the volume of the condensed phase, which we will return to momentarily. \todo{Do you need this point here? ** Note: I think that my other comments in this paragraph are addressed via panels b and c of Figure 3; perhaps those could be moved here.}

\begin{figure}[b!]
    \centerline{\includegraphics[scale=0.6]{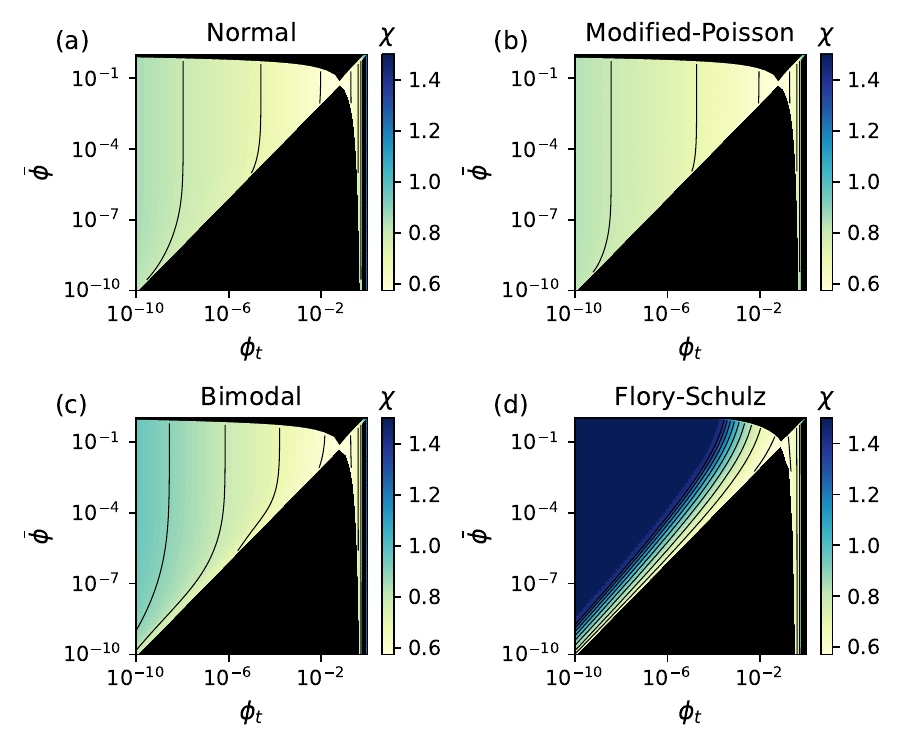}}
    %TC:ignore
    \caption{Full phase diagrams for different MW distributions. The overall phase diagram surface $\chi(\phi_t, \bar{\phi})$, where $\phi_t$ is the sum of all components in a given phase and $\bar{\phi}$ is the overall volume fraction of polymer added to the solution. The values are saturated at the bounds of the color bar.  The samples are (a) a normally distributed sample with $\langle N\rangle=200$ and $N_\sigma=20$, (b) a modified Poisson sample with $\lambda=200$, (c) a bimodal distribution with $\eta=0.5$, and (d) a Flory-Schulz sample with $b=0.99$. \color{black}The thin black lines correspond to constant-$\chi$ contours.\color{black} } %TC:endignore
    \label{Figure4}
    \end{figure}
        \begin{figure*}[t!]
\centerline{\includegraphics[scale=0.5]{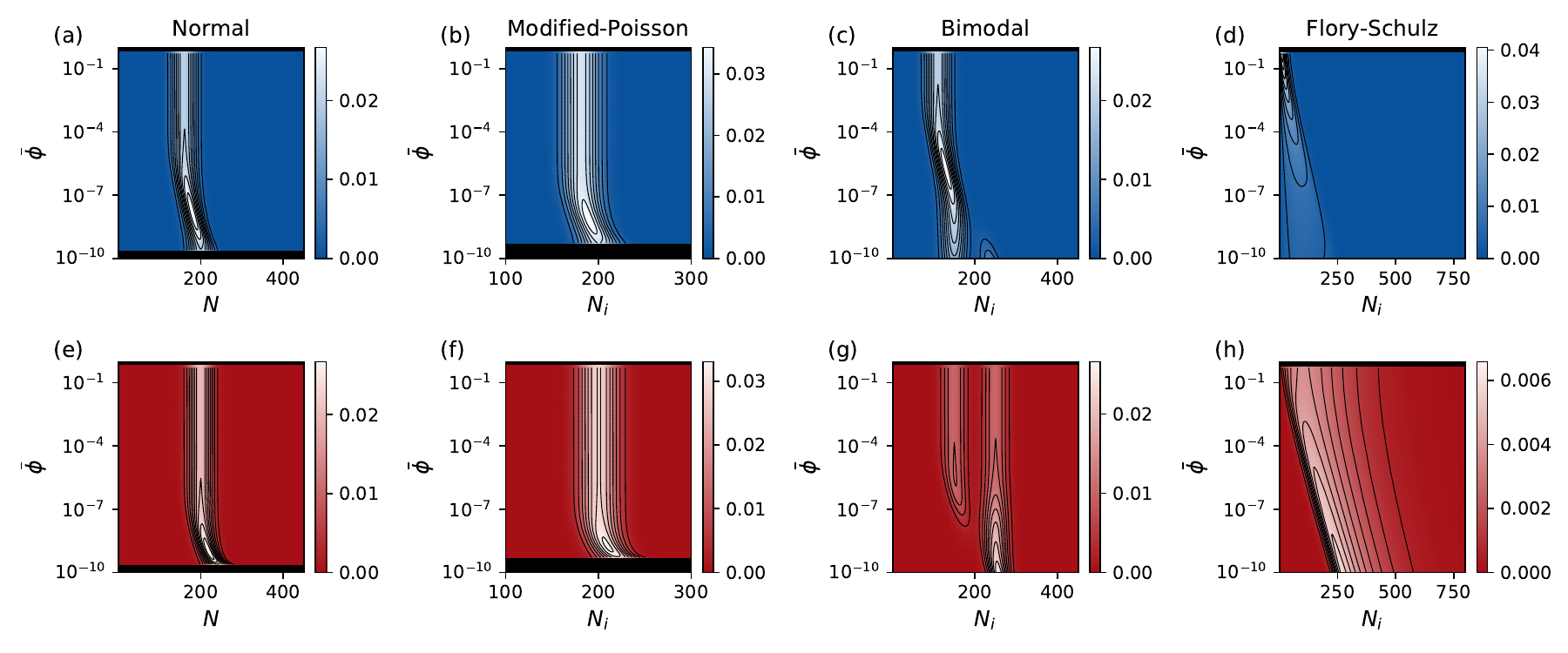}}
\caption{The MW distributions plotted as a function of the overall polymer volume fraction, $\bar{\phi}$, at a fixed value of $\chi=0.8$, corresponding to the phase diagrams in Figure~\ref{Figure4}. The top row (a-d) corresponds to the distribution in the dilute phase, while the bottom row (e-h) corresponds to the distribution in the concentrated phase. (a,e) correspond to the normally distributed sample with $<N>=200$ and $N_\sigma=20$, (b,f) correspond to the modified -Poisson distributed sample with $\lambda=200$, (c,g) correspond to the bimodal distributed sample with $\eta=0.5$, and (d,h) correspond to the Flory-Schulz distributed sample with $b=0.99$.   }
\label{Figure5}
\end{figure*}

\subsection{Coexistence curves}

Since the MW distribution in the condensed phase differs from the overall MW distribution, one may ask: is the coexistence curve itself sensitive to the MW distribution? This sensitivity has been found in previous studies of polydisperse distributions~\cite{sollich2001predicting}; here, we show how it can be easily quantified using our method. In Fig.~\ref{Figure3}, we  explore these features at the cloud point of the condensed phase in Fig.~\ref{Figure3}(a-d) (when the condensed phase first appears at $\nu=0$ with increasing polymer volume fraction) and the dilute phase in Fig.~\ref{Figure3}(e-h) (when the dilute phase first appears at $\nu=1$ with decreasing volume fraction) for the distributions from Table I by calculating the coexistence curves, expressed as $\chi(\phi_t)$. Note that each point along the coexistence curve corresponds to a unique MW distribution in each phase, but the phase diagram is simplified by projecting all phase compositions onto the coordinate $\phi_t$. In each plot, a thin dashed black line is plotted that matches the result if the polymer were purely monodisperse at the weight-average MW (Table I). For one distribution in each set, the cloud and shadow curves are plotted together in Fig.~\ref{Figure3}(i-l) with a linear scale on the $x$-axis. The cloud curve corresponds to $\nu=1$ on the condensed phase branch and to $\nu=0$ on the dilute phase branch. The corresponding shadow curve corresponds to $\nu=0$ on the condensed phase branch and to $\nu=1$ on the dilute phase branch.
% \todo{I was initially surprised that the binodal of the bimodal distribution does not depend as strongly on $\eta$ as that of the FS distribution depends on $b$, but then I realized that the bimodal distribution doesn't contain nearly as long polymers.  Is this the reason?  Could you put additional calculations in the SI?}

%TC:ignore

%TC:endignore

Clearly, the larger the polydispersity of the sample, the greater the shift of the coexistence curve from what one would expect for a monodisperse sample. This trend is most clearly seen for the normally distributed samples in Fig.~\ref{Figure3}(a,e) with varying standard deviation. Further, the degree of shifting in the coexistence curve becomes more significant at larger values of $\chi$, as the phase equilibrium becomes more dominated by the larger MW components. These shifts can be significant, considering that the $\phi_t$ axis is plotted with a logarithmic scale. 

There are also cases where the coexistence curve is less sensitive to the MW distribution. For example, the modified-Poisson distributed samples in Fig.~\ref{Figure3}(b,f) exhibit a weak shifting of the curve, which is only relevant at large $\chi$. Regardless of the ratio between peaks, the bimodal distributions in Fig.~\ref{Figure3}(c,g) are dominated by the higher MW peak at $\nu=0$ and by the lower peak at $\nu=1$ regardless of the ratio between the peak heights. The largest deviations from the monodisperse behavior are found for the Flory-Schulz distribution in Fig.~\ref{Figure3}(d,h), which is consistent with the extreme fractionation in these distributions with longer tails.

These coexistence curves at infinitesimal condensed phase, $\nu=0$, and infinitesimal dilute phase $\nu=1$, can be used, respectively, to construct the cloud and shadow curves~\cite{sollich2001predicting} of the polydisperse systems in Fig.~\ref{Figure3}(i-j). For all the distributions tested, there is a misalignment of the cloud and shadow curves owing to the fractionation of the polymers between the phases. The misalignment of these curves gives an indication of the extent of fractionation.

% {\color{red} do you do this next?}

% \todo{I'm having difficulty rationalizing the differences between the FS and bimodal results.  We should discuss this point when we meet!}

\begin{figure}[h!]
\centerline{\includegraphics[scale=0.3]{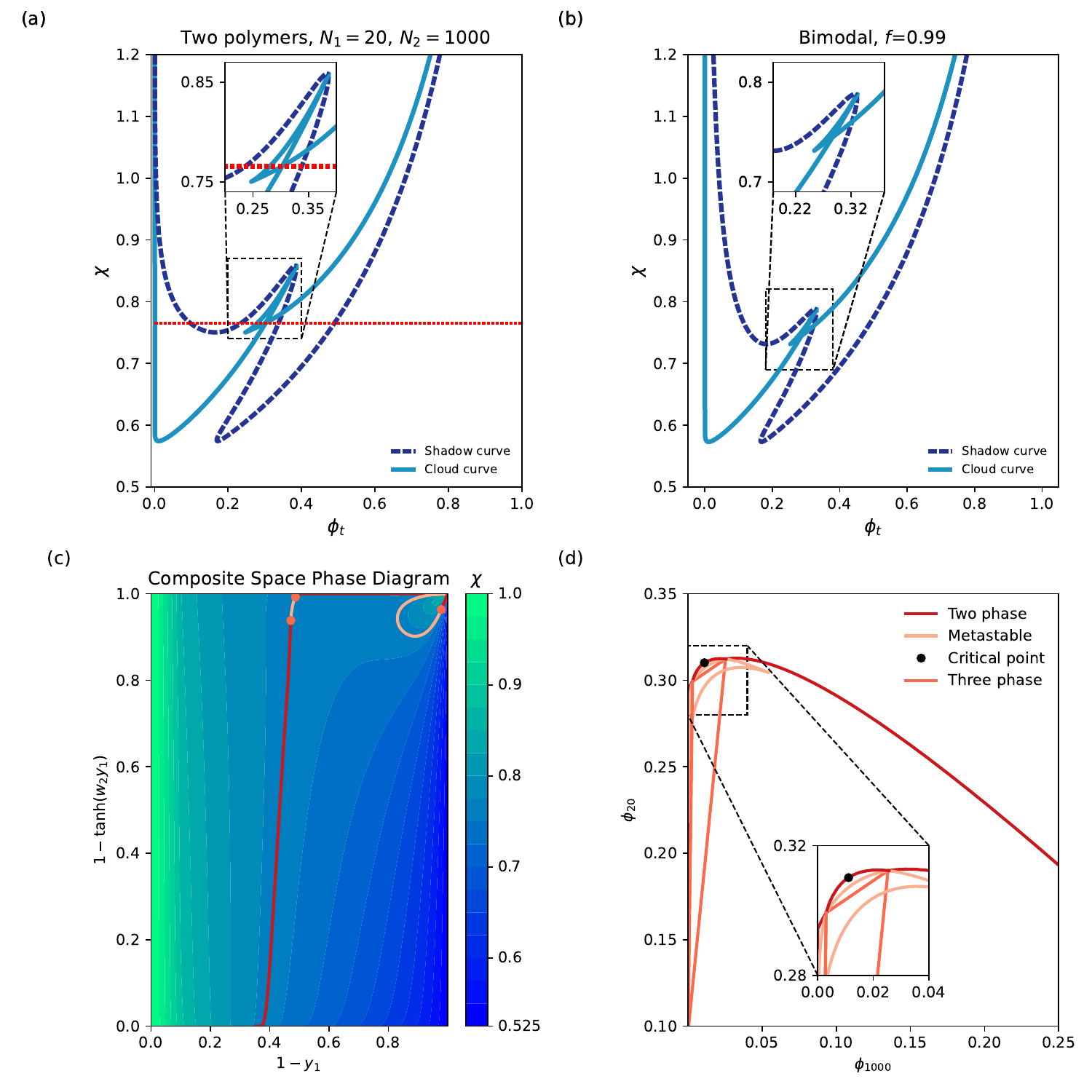}}
\caption{Three-phase coexistence for polymers of disparate lengths. The cloud and shadow curve for a system of two polymers (a) with $N_1=20$ and $N_2=1000$ with 1\% high MW fraction or (b)a bimodal distributed sample with centers $N_1=20$ and $N_2=1000$ again with standard deviation $N_\sigma=20$ and $\eta=0.99$. The dotted red line in $a$ corresponds to the value of $\chi=0.76555$ used for calculations in (d). (c) For the two polymer system in (a), the full composite space phase diagram is plotted as a $\chi$ surface in terms of $1-y_1$ and $1-\tanh(w_2 y_1)$. The color bar corresponds to value of $\chi$. The red curve corresponds to the coexistence at $\chi=0.76555$. The orange markers correspond to the three-phase coexistence points. The light pink curves correspond to the metastable binodal points that are within the two-phase or three-phase coexistence region. (d) The real-composition phase diagram at fixed $\chi=0.76555$. the line colors are identical to (c) indicating two and three-phase coexistence and metastable binodals. The points within the orange triangle correspond to the three-phase coexistence region.   }
\label{FigureS4}
\end{figure}

\subsection{Full phase diagrams}
So far, we have presented results at specified values of $\nu=0$ and $\nu=1$. Yet, our method can be efficiently applied to the full two-phase composition regime over all values of $\nu$. We turn to the calculation of the overall phase diagram in Fig.~\ref{Figure4}. In Fig.~\ref{Figure4}(a), the overall phase diagram is projected to the surface $\chi(\phi_t,\bar{\phi})$, where $\phi_t$ corresponds to the total polymer volume fraction in each phase, $\phi_{tB}$ and $\phi_{tA}$, and $\bar{\phi}$ corresponds to the overall volume fraction of polymer added to solvent.  This surface plot embeds a lot of information, but it can be read as follows: (i) Draw a horizontal line corresponding to the overall volume fraction of polymer, $\bar{\phi}$. (ii) Find the intersection of the horizontal line with the corresponding value of $\chi$ on the left and right colored regions. The intersection in the left colored region gives the dilute phase $\phi_{tB}$, while the intersection on the right colored region gives the condensed phase volume fraction $\phi_{tA}$. The point where the two regions meet is the global critical point.

% {\color{red} HAS: earlier when you refer to $\phi_t$, such as when calculating the cloud point you are really referring to $\bar{\phi}$ (or $\psi_t$ in the dilute phase?))?}

For monodisperse samples at fixed $\chi$, the dilute and condensed phases have fixed composition regardless of the overall volume fraction of the polymer. Therefore, an ideal monodisperse sample would have contours on the $\chi$ surface that are perfectly vertical, which is closely mirrored by the modified-Poisson sample displayed in Fig.~\ref{Figure4}(b). On the other hand, a shift in the bulk composition of the two phases with increasing overall polymer volume fraction corresponds to angled contours on the $\chi$ surface. The normally distributed sample exhibits some of this angled behavior in Fig.~\ref{Figure4}(a), which is comparatively more extreme for the bimodal and Flory-Schulz distributions in Fig.~\ref{Figure4}(c) and(d), respectively. The Flory-Schulz distribution again exhibits the greatest deviation relative to monodisperse behavior. The $\chi$-contours appear to be nearly at 45$^\circ$ on the dilute side of the $\chi(\phi_t, \bar{\phi})$ surface.

% \todo{It would be helpful if we could see a direct comparison with the parent/dry distribution in these panels.  Also, see my comment about moving these parts earlier in the presentation.} The corresponding MW distributions for the other types of distributions at a sample value of $\chi=0.8$ are given in the SI.

Each point on the phase diagram corresponds to a unique MW distribution in the condensed and dilute phases. If we follow the $\chi=0.8$ contour on Fig.~\ref{Figure4}(a) as $\bar{\phi}$ is increased, we can directly evaluate how the MW distribution changes while increasing the overall polymer volume fraction. In  Fig.~\ref{Figure5}, the MW distribution is plotted as $p(N; \bar{\phi})$ for the dilute and condensed phases, respectively. As the polymer volume fraction is increased, the condensed phase (A) distribution eventually matches the overall MW distribution. Conversely, the dilute phase (B) distribution goes from matching the overall distribution to selecting for shorter MW components as $\bar{\phi}$ increases. 
% {\color{red} to my reading this last sentence contradicts the discussion back in figures 1 and 2.}

\subsection{Three-phase coexistence}
It was proved early on theoretically and experimentally \cite{tompa1949phase, koningsveld2001polymer} that polydispersity in length can lead to three-phase coexistence even for homopolymers, albeit over narrow temperature ranges. This phenomenon is especially notable since the modulation of entropy of chains is enough to segregate chains in multiple phases even without differences in enthalpic interactions between monomers. While monomer interactions are indistinguishable, the interactions per chain depend on the polymer length, leading to differential segregation into three phases.   With the previous results presented on two-phase coexistence, we asked: Can the method that is designed for analytical two-phase coexistence calculations apply to these small but physically relevant mutli-phase regions?

To answer this question, in Fig. 6(a), we calculated the cloud and shadow curves for a system of a polymer containing two chain lengths, similar to phase diagrams presented in $~\cite{koningsveld2001polymer}$ using the same two-phase calculation method described previously. The cloud and shadow curves display striking non-monotonicity corresponding to multiple equilibrium solutions to the two-phase coexistence problem. For example, the cloud curve loops back on itself, with two shadow curve points pointing back to the same cloud curve point at the cloud curve intersection. Within the cloud point loop, there are three cloud curve points and three shadow curve points at a given fixed value of $\chi$. The presence of a three-phase region is therefore indicated by the intersection in the exact calculation of the cloud and shadow curves using the implicit substitution method for two phases. The complexity of the cloud and shadow curves points is maintained even if polydispersity is present in the two polymer fractions. In Fig. 6(b), a bimodal distribution with the same relative fractions as in  Fig. 6(a) is added but now with finite variance, $N_\sigma$, and the cloud curve maintains a similar looped structure.

We can also find these three-phase coexistence regions in the transformed coordinate space that we use within the implicit substitution method, as shown in Fig 6(c), evaluated over all possible values with two degrees of freedom in composite coordinates for the polymer mixture system from Fig. 6(a). The appearance of more than one binodal solution to two-phase coexistence appears as multiple disconnected contours in the transformed space, as shown with the highlighted contour in red and pink at $\chi=0.76555$. However without transforming to real composition space, it is not possible to determine which points correspond to the two-phase or three-phase region, or which points are metastable binodal points~\cite{koningsveld2001polymer}. We can transform the contour from Fig. 6(c) to the coordinates in  Fig. 6(d). Based on the intersection of tie-lines, we can identify the relevant multiphase regions, which are color-coded in the plot as described in the figure caption. 

Clearly, the appearance of three phases may complicate the application of the two-phase calculations employed in this work. Finding the three-phase compositions at all temperatures requires direct mapping from composite space to real space and the identification of intersections of two-phase binodal curves. However, the efficient calculation of the two-phase cloud and shadow curves can be used to infer the behavior and locations of the three-phase regions, which provides a jump start on this challenging computational task for arbitrary mixtures.

\section{Conclusions}

In summary, by direct evaluation of analytical formulas of polymer fractionation, we have shown that the shape of the MW distribution is crucial  in phase coexistence calculations. The greater the polydispersity, the greater the fractionation of MWs between phases and the shift of the coexistence curve relative to monodisperse samples. The tail of the MW distribution dominates in the phase coexistence calculations, especially for distributions with long tails or for bimodal distributions near their cloud point. Three phase coexistence in polydisperse mixtures requires more care, but three-phase coexistence regions are captured by the method when they occur. These findings are consistent with early studies~\cite{sollich2001predicting}, but provide an efficient route to fast and adaptable calculations.

The approaches presented here may allow for an interpretation of the phase behavior of polymer samples including consideration of a realistic distribution of MWs with straightforward computations. Further, our analysis may be useful for designing a polymer fractionation process by varying solvent conditions. To allow for ease of phase coexistence calculations with polymer polydispersity, we have provided a python code~\cite{GitCode} that allows for two-phase calculations with an arbitrary MW distribution. 

In our analysis, we have performed all calculations in terms of $\chi$, but the exact value of $\chi$ may depend on temperature and the solvent identity. The calculations have been performed assuming that all polymers have the same monomer composition, and future work could extend the analysis to copolymer systems with polydispersity in size and chemical composition~\cite{van2023quantitative, chen2023emergence}. Further, while the calculations done here correspond to a \textit{batch} process in which there are no spatial gradients or temporal evolution in composition, fractionation processes may be designed in which the overall MW distribution (and in each phase) may become a function of space and time\cite{francuskiewicz2013polymer}. 

The analytical approach explored here may be extended to such scenarios where model-based simulation and optimization would be aided by efficient phase coexistence calculations. The approach allows for the computationally-efficient consideration of the entire MW distribution at high numerical accuracy, providing a method to describe the essential role of polydispersity in polymeric phase separation.

\begin{acknowledgments}

%TC:ignore
JPD and HAS acknowledge support from the Princeton Center for Complex Materials, an NSF MRSEC (DMR-2011750), and NSF for grant DMS/NIGMS 2245850. JPD is also supported by the Omenn-Darling Bioengineering Institute -- Innovators (ODBI2) Postdoctoral Fellowship. WMJ acknowledges support from the National Institute of General Medical Sciences of the National Institutes of Health under award number R35GM155017.
%TC:endignore

\end{acknowledgments}

% %%%%%%%%%%%%%%%%%%%%%%%%%%%%%%%%%%%%%%%%%%%%%%%%%%%%%%%%%%%%%%%%%%%%%
% %% The same is true for Supporting Information, which should use the
% %% suppinfo environment.
% %%%%%%%%%%%%%%%%%%%%%%%%%%%%%%%%%%%%%%%%%%%%%%%%%%%%%%%%%%%%%%%%%%%%%
% \section*{Supplemental Material}

% %TC:ignore
% Detailed description of fractionation calculations, discussion of previous fractionation calculation approaches, and supplementary plots characterizing phase diagrams.
% %TC:endignore

\appendix
\section{Fractionation at the tails of the distribution at the cloud point}
Here, we will analyze fractionation at the tails of the distribution far from the critical point, but located at the cloud point, $\nu=0$. In this limit, we can safely assume that $y(N_i)\approx1-\epsilon(N_i)$ where $\epsilon(N_i)\ll 1$. In this limit, with $\epsilon_1=\epsilon(N_1)$, using equation (\ref{eq:eqyN}), we get
\begin{equation}
    \epsilon(N_i)=2\left(\frac{\epsilon_1}{2}\right)^{N_i/N_1}.
\end{equation}
The volume fraction of component with length $N_i$ relative to the mass average length, $N_1$, in the condensed phase is defined by the ratio
\begin{equation}
    \frac{\phi_A(N_i)}{\phi_A(N_1)}=\frac{\beta(N_i)(1+y(N_i))}{\beta(N_1)(1+y(N_1))}\approx{w(N_i)}
\end{equation}
with $w(N_1)=1$. At the cloud point, $\nu=0$, and with equation (\ref{eq:eqwN}), we can relate $w(N_i)$ to $\epsilon_1$, 
\begin{equation} \label{eq:eqWasymptotic}
    \frac{\phi_A(N_i)}{\phi_A(N_1)}\approx \frac{p(N_i)}{p(N_1)}\left(\frac{\epsilon_1}{2}\right)^{1-N_i/N_1}
\end{equation}
Equation (\ref{eq:eqWasymptotic}) highlights the asymptotic behavior far from the critical point but also at the cloud point. If the MW distribution decays more slowly than exponential, then the  highest molecular weight species will dominate the condensed phase, as was seen for the Flory-Schulz MW distribution.

By evaluating the weighting functions in this limit, we can determine the limiting molecular weight distribution far from the critical point. The weighting functions take on the limiting forms
\begin{subequations}
    \begin{equation}
    Q_1(N_i)=\left(\frac{\epsilon_1}{2}\right)^{1-N_i/N_1}
    % Q_1(N_i)\approx\frac{\nu +\epsilon_1/2}{\nu+(\epsilon_1/2)^{N/N_1}}\begin{cases} (\epsilon_1/2)^{(N_1-N)/N_1}&\nu=0 \\ 
    % 1+\frac{\epsilon_1}{2\nu}-\frac{1}{\nu}(\frac{\epsilon_1}{2})^{N/N_1} &\nu\neq 0
    % \end{cases}
\end{equation}
\begin{equation}
    Q_2(N_i)\approx \left(-\frac{1}{2N_1}\ln(\epsilon_1/2)-\frac{1}{N}\right)Q_1(N_i),
\end{equation}
\begin{equation}
    Q_3(N_i)\approx (1+2(\epsilon_1/2)^{N/N_1})Q_1(N_i)
\end{equation}
and the partioning of each component is
\begin{equation}
    y(N_i)\approx 1-2(\epsilon_1/2)^{N/N_1}.
\end{equation}
\end{subequations}

In terms of the $N_i$ dependence of the distribution, we find the following relationship far from the critical point for the condensed and dilute phases, respectively,
\begin{subequations}
    \begin{equation}
        \phi_A(N_i)\propto p(N_i) \left(\frac{\epsilon_1}{2}\right)^{1-N_i/N_1}
    \end{equation}
    \begin{equation}
        \phi_B(N_i)\propto p(N_i) \epsilon_1.
    \end{equation}
\end{subequations}

\section{Inverse FH function calculation}
Here, we pursue a series representation of the inverse Flory-Huggins (FH) function, which is related to the Generalized-LambertW function. 
The FH function has the form
\begin{equation}
    h(z)=\frac{\tanh^{-1}(z)}{z}=\frac{1}{2z}\ln\left(\frac{1+z}{1-z}\right)=A,
\end{equation}

First, we define the variable $x=-2 z A$, so that the equation takes the convenient form
\begin{equation}
    2A-x=\left(2A+ x\right)e^{-x},
\end{equation}
or 
\begin{equation}
    x=-2A+(2A-x)e^x.
\end{equation}
Using the Lagrange inversion theorem~\cite{NIST:DLMF:1.10.vii}, the solution for $x$ can be described by:
\begin{equation}
    x=-2A+\sum_{k=1}^\infty\frac{1}{k!}\left(\frac{d}{d x }\right)^{k-1}\left[\left(2A- x\right)^k e^{k x}\right]_{x=-2A}.
\end{equation}
We can then define $u=-kx+2Ak$
\begin{equation}
    x=-2A+\sum_{k=1}^\infty\frac{(-k)^{k-1}}{k!}\left(\frac{d}{d u }\right)^{k-1}\left[\left(\frac{u}{k}\right)^k e^{-u+2Ak}\right]_{u=4Ak}
\end{equation}
and reformulate as:
\begin{equation}
    x=-2A-\sum_{k=1}^\infty\frac{(-1)^{k}}{k!k}e^{2Ak}\left(\frac{d}{d u }\right)^{k-1}\left[u^k e^{-u}\right]_{u=4Ak}.
\end{equation}
Next, we make use of the following identity
\begin{equation}
    \left(\frac{d}{d u }\right)^{k-1}\left[u^k e^{-u}\right]=(k-1)! u e^{-u}L_{k-1}^{(1)}(u)
\end{equation}
where $L()$ signifies the generalized Laguerre polynomials, so that we can write the final inversion formula as
\begin{equation}
    x=-2A-\sum_{k=1}^\infty\frac{4A(-1)^{k}}{k}e^{-2Ak}L_{k-1}^{(1)}(4Ak),
\end{equation}
or in terms of $z$, we get:
\begin{equation}
    z=1+2\sum_{k=1}^\infty\frac{(-1)^{k}}{k}e^{-2Ak}L_{k-1}^{(1)}(4Ak).
\end{equation}

The series converges well for intermediate values of $1.076<A<5$, which we can confirm by evaluating 100 terms in the series. To maintain accuracy outside of these bounds (since $A$ can vary between 1 and $\infty$), we apply asymptotic solutions for $A$ in the limit of large and small $A$. For $A<1.076$, we expand up to sixth order and tuncate,
\begin{equation}
    h(z)\approx 1+\frac{z^2}{3}+\frac{z^4}{5}+\frac{z^6}{7}=A,
\end{equation}
then solve a cubic equation for $z^2$, keeping the real, positive root of $z$. 

For $A>5$, we use the limiting formula:
\begin{equation}
    h(z)\approx \frac{1}{2}\ln\left(\frac{2}{1-z}\right)=A
\end{equation}
from which we can derive:
\begin{equation}
    z=1-2e^{-2A},
\end{equation}
valid at large values of $A$. 

Using these series and asymptotic expressions, the maximum error for
\begin{equation}
    h\left(h^{-1}(A)\right)
\end{equation}
is within 0.02\% of $A$ over the entire possible range of $A$.

\bibliographystyle{apsrev4-2}
\bibliography{library}

\begin{thebibliography}{44}%
\makeatletter
\providecommand \@ifxundefined [1]{%
 \@ifx{#1\undefined}
}%
\providecommand \@ifnum [1]{%
 \ifnum #1\expandafter \@firstoftwo
 \else \expandafter \@secondoftwo
 \fi
}%
\providecommand \@ifx [1]{%
 \ifx #1\expandafter \@firstoftwo
 \else \expandafter \@secondoftwo
 \fi
}%
\providecommand \natexlab [1]{#1}%
\providecommand \enquote  [1]{``#1''}%
\providecommand \bibnamefont  [1]{#1}%
\providecommand \bibfnamefont [1]{#1}%
\providecommand \citenamefont [1]{#1}%
\providecommand \href@noop [0]{\@secondoftwo}%
\providecommand \href [0]{\begingroup \@sanitize@url \@href}%
\providecommand \@href[1]{\@@startlink{#1}\@@href}%
\providecommand \@@href[1]{\endgroup#1\@@endlink}%
\providecommand \@sanitize@url [0]{\catcode `\\12\catcode `\$12\catcode `\&12\catcode `\#12\catcode `\^12\catcode `\_12\catcode `\%12\relax}%
\providecommand \@@startlink[1]{}%
\providecommand \@@endlink[0]{}%
\providecommand \url  [0]{\begingroup\@sanitize@url \@url }%
\providecommand \@url [1]{\endgroup\@href {#1}{\urlprefix }}%
\providecommand \urlprefix  [0]{URL }%
\providecommand \Eprint [0]{\href }%
\providecommand \doibase [0]{https://doi.org/}%
\providecommand \selectlanguage [0]{\@gobble}%
\providecommand \bibinfo  [0]{\@secondoftwo}%
\providecommand \bibfield  [0]{\@secondoftwo}%
\providecommand \translation [1]{[#1]}%
\providecommand \BibitemOpen [0]{}%
\providecommand \bibitemStop [0]{}%
\providecommand \bibitemNoStop [0]{.\EOS\space}%
\providecommand \EOS [0]{\spacefactor3000\relax}%
\providecommand \BibitemShut  [1]{\csname bibitem#1\endcsname}%
\let\auto@bib@innerbib\@empty
%</preamble>
\bibitem [{\citenamefont {Flory}(1953)}]{flory1953principles}%
  \BibitemOpen
  \bibfield  {author} {\bibinfo {author} {\bibfnamefont {P.~J.}\ \bibnamefont {Flory}},\ }\href@noop {} {\emph {\bibinfo {title} {Principles of Polymer Chemistry}}}\ (\bibinfo  {publisher} {Cornell University Press},\ \bibinfo {year} {1953})\BibitemShut {NoStop}%
\bibitem [{\citenamefont {Cantow}(1967)}]{cantow2013polymer}%
  \BibitemOpen
  \bibfield  {author} {\bibinfo {author} {\bibfnamefont {M.~J.}\ \bibnamefont {Cantow}},\ }\href@noop {} {\emph {\bibinfo {title} {Polymer Fractionation}}}\ (\bibinfo  {publisher} {Academic Press Inc},\ \bibinfo {year} {1967})\BibitemShut {NoStop}%
\bibitem [{\citenamefont {Francuskiewicz}(2013)}]{francuskiewicz2013polymer}%
  \BibitemOpen
  \bibfield  {author} {\bibinfo {author} {\bibfnamefont {F.}~\bibnamefont {Francuskiewicz}},\ }\href@noop {} {\emph {\bibinfo {title} {Polymer Fractionation}}}\ (\bibinfo  {publisher} {Springer Science \& Business Media},\ \bibinfo {year} {2013})\BibitemShut {NoStop}%
\bibitem [{\citenamefont {Hiemenz}\ and\ \citenamefont {Lodge}(2007)}]{hiemenz2007polymer}%
  \BibitemOpen
  \bibfield  {author} {\bibinfo {author} {\bibfnamefont {P.~C.}\ \bibnamefont {Hiemenz}}\ and\ \bibinfo {author} {\bibfnamefont {T.~P.}\ \bibnamefont {Lodge}},\ }\href@noop {} {\emph {\bibinfo {title} {Polymer Chemistry}}}\ (\bibinfo  {publisher} {CRC Press},\ \bibinfo {year} {2007})\BibitemShut {NoStop}%
\bibitem [{\citenamefont {Gentekos}\ \emph {et~al.}(2019)\citenamefont {Gentekos}, \citenamefont {Sifri},\ and\ \citenamefont {Fors}}]{gentekos2019controlling}%
  \BibitemOpen
  \bibfield  {author} {\bibinfo {author} {\bibfnamefont {D.~T.}\ \bibnamefont {Gentekos}}, \bibinfo {author} {\bibfnamefont {R.~J.}\ \bibnamefont {Sifri}},\ and\ \bibinfo {author} {\bibfnamefont {B.~P.}\ \bibnamefont {Fors}},\ }\href@noop {} {\bibfield  {journal} {\bibinfo  {journal} {Nature Reviews Materials}\ }\textbf {\bibinfo {volume} {4}},\ \bibinfo {pages} {761} (\bibinfo {year} {2019})}\BibitemShut {NoStop}%
\bibitem [{\citenamefont {Odian}(2004)}]{odian2004principles}%
  \BibitemOpen
  \bibfield  {author} {\bibinfo {author} {\bibfnamefont {G.}~\bibnamefont {Odian}},\ }\href@noop {} {\emph {\bibinfo {title} {Principles of Polymerization}}}\ (\bibinfo  {publisher} {John Wiley \& Sons},\ \bibinfo {year} {2004})\BibitemShut {NoStop}%
\bibitem [{\citenamefont {Billmeyer~Jr.}(1965)}]{billmeyer1965characterization}%
  \BibitemOpen
  \bibfield  {author} {\bibinfo {author} {\bibfnamefont {F.~W.}\ \bibnamefont {Billmeyer~Jr.}},\ }\href@noop {} {\bibfield  {journal} {\bibinfo  {journal} {Journal of Polymer Science Part C: Polymer Symposia}\ }\textbf {\bibinfo {volume} {8}},\ \bibinfo {pages} {161} (\bibinfo {year} {1965})}\BibitemShut {NoStop}%
\bibitem [{\citenamefont {Whitfield}\ \emph {et~al.}(2019)\citenamefont {Whitfield}, \citenamefont {Truong}, \citenamefont {Messmer}, \citenamefont {Parkatzidis}, \citenamefont {Rolland},\ and\ \citenamefont {Anastasaki}}]{whitfield2019tailoring}%
  \BibitemOpen
  \bibfield  {author} {\bibinfo {author} {\bibfnamefont {R.}~\bibnamefont {Whitfield}}, \bibinfo {author} {\bibfnamefont {N.~P.}\ \bibnamefont {Truong}}, \bibinfo {author} {\bibfnamefont {D.}~\bibnamefont {Messmer}}, \bibinfo {author} {\bibfnamefont {K.}~\bibnamefont {Parkatzidis}}, \bibinfo {author} {\bibfnamefont {M.}~\bibnamefont {Rolland}},\ and\ \bibinfo {author} {\bibfnamefont {A.}~\bibnamefont {Anastasaki}},\ }\href@noop {} {\bibfield  {journal} {\bibinfo  {journal} {Chemical Science}\ }\textbf {\bibinfo {volume} {10}},\ \bibinfo {pages} {8724} (\bibinfo {year} {2019})}\BibitemShut {NoStop}%
\bibitem [{\citenamefont {Mencer}(1988)}]{mencer1988efficiency}%
  \BibitemOpen
  \bibfield  {author} {\bibinfo {author} {\bibfnamefont {H.}~\bibnamefont {Mencer}},\ }\href@noop {} {\bibfield  {journal} {\bibinfo  {journal} {Polymer Engineering \& Science}\ }\textbf {\bibinfo {volume} {28}},\ \bibinfo {pages} {497} (\bibinfo {year} {1988})}\BibitemShut {NoStop}%
\bibitem [{\citenamefont {Wang}\ \emph {et~al.}(2023)\citenamefont {Wang}, \citenamefont {Luo},\ and\ \citenamefont {Zhou}}]{wang2023precise}%
  \BibitemOpen
  \bibfield  {author} {\bibinfo {author} {\bibfnamefont {T.-T.}\ \bibnamefont {Wang}}, \bibinfo {author} {\bibfnamefont {Z.-H.}\ \bibnamefont {Luo}},\ and\ \bibinfo {author} {\bibfnamefont {Y.-N.}\ \bibnamefont {Zhou}},\ }\href@noop {} {\bibfield  {journal} {\bibinfo  {journal} {Macromolecules}\ }\textbf {\bibinfo {volume} {56}},\ \bibinfo {pages} {1130} (\bibinfo {year} {2023})}\BibitemShut {NoStop}%
\bibitem [{\citenamefont {van Leuken}\ \emph {et~al.}(2023{\natexlab{a}})\citenamefont {van Leuken}, \citenamefont {van Osch}, \citenamefont {Kouris}, \citenamefont {Yao}, \citenamefont {Jedrzejczyk}, \citenamefont {Cremers}, \citenamefont {Bernaerts}, \citenamefont {van Benthem}, \citenamefont {Tuinier}, \citenamefont {Boot}, \citenamefont {Hensen},\ and\ \citenamefont {Vis}}]{van2023quantitative}%
  \BibitemOpen
  \bibfield  {author} {\bibinfo {author} {\bibfnamefont {S.~H.~M.}\ \bibnamefont {van Leuken}}, \bibinfo {author} {\bibfnamefont {D.~J. G.~P.}\ \bibnamefont {van Osch}}, \bibinfo {author} {\bibfnamefont {P.~D.}\ \bibnamefont {Kouris}}, \bibinfo {author} {\bibfnamefont {Y.}~\bibnamefont {Yao}}, \bibinfo {author} {\bibfnamefont {M.~A.}\ \bibnamefont {Jedrzejczyk}}, \bibinfo {author} {\bibfnamefont {G.~J.~W.}\ \bibnamefont {Cremers}}, \bibinfo {author} {\bibfnamefont {K.~V.}\ \bibnamefont {Bernaerts}}, \bibinfo {author} {\bibfnamefont {R.~A. T.~M.}\ \bibnamefont {van Benthem}}, \bibinfo {author} {\bibfnamefont {R.}~\bibnamefont {Tuinier}}, \bibinfo {author} {\bibfnamefont {M.~D.}\ \bibnamefont {Boot}}, \bibinfo {author} {\bibfnamefont {E.~J.~M.}\ \bibnamefont {Hensen}},\ and\ \bibinfo {author} {\bibfnamefont {M.}~\bibnamefont {Vis}},\ }\href {https://doi.org/10.1039/D3GC00948C} {\bibfield  {journal} {\bibinfo  {journal} {Green Chem.}\ }\textbf {\bibinfo {volume} {25}},\ \bibinfo {pages} {7534} (\bibinfo {year}
  {2023}{\natexlab{a}})}\BibitemShut {NoStop}%
\bibitem [{\citenamefont {Ward}\ and\ \citenamefont {Sweeney}(2012)}]{ward2012mechanical}%
  \BibitemOpen
  \bibfield  {author} {\bibinfo {author} {\bibfnamefont {I.~M.}\ \bibnamefont {Ward}}\ and\ \bibinfo {author} {\bibfnamefont {J.}~\bibnamefont {Sweeney}},\ }\href@noop {} {\emph {\bibinfo {title} {Mechanical Properties of Solid Polymers}}}\ (\bibinfo  {publisher} {John Wiley \& Sons},\ \bibinfo {year} {2012})\BibitemShut {NoStop}%
\bibitem [{\citenamefont {Bird}\ \emph {et~al.}(1987)\citenamefont {Bird}, \citenamefont {Armstrong},\ and\ \citenamefont {Hassager}}]{bird1987dynamics}%
  \BibitemOpen
  \bibfield  {author} {\bibinfo {author} {\bibfnamefont {R.~B.}\ \bibnamefont {Bird}}, \bibinfo {author} {\bibfnamefont {R.~C.}\ \bibnamefont {Armstrong}},\ and\ \bibinfo {author} {\bibfnamefont {O.}~\bibnamefont {Hassager}},\ }\href@noop {} {\emph {\bibinfo {title} {Dynamics of Polymeric Liquids. Vol. 1: Fluid Mechanics}}}\ (\bibinfo  {publisher} {John Wiley and Sons Inc., New York, NY},\ \bibinfo {year} {1987})\BibitemShut {NoStop}%
\bibitem [{\citenamefont {Godovsky}(2012)}]{godovsky2012thermophysical}%
  \BibitemOpen
  \bibfield  {author} {\bibinfo {author} {\bibfnamefont {Y.~K.}\ \bibnamefont {Godovsky}},\ }\href@noop {} {\emph {\bibinfo {title} {Thermophysical Properties of Polymers}}}\ (\bibinfo  {publisher} {Springer Science \& Business Media},\ \bibinfo {year} {2012})\BibitemShut {NoStop}%
\bibitem [{\citenamefont {Ward}\ and\ \citenamefont {Georgiou}(2011)}]{ward2011thermoresponsive}%
  \BibitemOpen
  \bibfield  {author} {\bibinfo {author} {\bibfnamefont {M.~A.}\ \bibnamefont {Ward}}\ and\ \bibinfo {author} {\bibfnamefont {T.~K.}\ \bibnamefont {Georgiou}},\ }\href@noop {} {\bibfield  {journal} {\bibinfo  {journal} {Polymers}\ }\textbf {\bibinfo {volume} {3}},\ \bibinfo {pages} {1215} (\bibinfo {year} {2011})}\BibitemShut {NoStop}%
\bibitem [{\citenamefont {Blythe}\ and\ \citenamefont {Bloor}(2005)}]{blythe2005electrical}%
  \BibitemOpen
  \bibfield  {author} {\bibinfo {author} {\bibfnamefont {A.~R.}\ \bibnamefont {Blythe}}\ and\ \bibinfo {author} {\bibfnamefont {D.}~\bibnamefont {Bloor}},\ }\href@noop {} {\emph {\bibinfo {title} {Electrical Properties of Polymers}}}\ (\bibinfo  {publisher} {Cambridge University Press},\ \bibinfo {year} {2005})\BibitemShut {NoStop}%
\bibitem [{\citenamefont {Higashihara}\ and\ \citenamefont {Ueda}(2015)}]{higashihara2015recent}%
  \BibitemOpen
  \bibfield  {author} {\bibinfo {author} {\bibfnamefont {T.}~\bibnamefont {Higashihara}}\ and\ \bibinfo {author} {\bibfnamefont {M.}~\bibnamefont {Ueda}},\ }\href@noop {} {\bibfield  {journal} {\bibinfo  {journal} {Macromolecules}\ }\textbf {\bibinfo {volume} {48}},\ \bibinfo {pages} {1915} (\bibinfo {year} {2015})}\BibitemShut {NoStop}%
\bibitem [{\citenamefont {Shin}\ and\ \citenamefont {Brangwynne}(2017)}]{shin2017liquid}%
  \BibitemOpen
  \bibfield  {author} {\bibinfo {author} {\bibfnamefont {Y.}~\bibnamefont {Shin}}\ and\ \bibinfo {author} {\bibfnamefont {C.~P.}\ \bibnamefont {Brangwynne}},\ }\href@noop {} {\bibfield  {journal} {\bibinfo  {journal} {Science}\ }\textbf {\bibinfo {volume} {357}},\ \bibinfo {pages} {eaaf4382} (\bibinfo {year} {2017})}\BibitemShut {NoStop}%
\bibitem [{\citenamefont {Banani}\ \emph {et~al.}(2017)\citenamefont {Banani}, \citenamefont {Lee}, \citenamefont {Hyman},\ and\ \citenamefont {Rosen}}]{banani2017biomolecular}%
  \BibitemOpen
  \bibfield  {author} {\bibinfo {author} {\bibfnamefont {S.~F.}\ \bibnamefont {Banani}}, \bibinfo {author} {\bibfnamefont {H.~O.}\ \bibnamefont {Lee}}, \bibinfo {author} {\bibfnamefont {A.~A.}\ \bibnamefont {Hyman}},\ and\ \bibinfo {author} {\bibfnamefont {M.~K.}\ \bibnamefont {Rosen}},\ }\href@noop {} {\bibfield  {journal} {\bibinfo  {journal} {Nature Reviews Molecular Cell Biology}\ }\textbf {\bibinfo {volume} {18}},\ \bibinfo {pages} {285} (\bibinfo {year} {2017})}\BibitemShut {NoStop}%
\bibitem [{\citenamefont {Jacobs}(2023)}]{jacobs2023theory}%
  \BibitemOpen
  \bibfield  {author} {\bibinfo {author} {\bibfnamefont {W.~M.}\ \bibnamefont {Jacobs}},\ }\href@noop {} {\bibfield  {journal} {\bibinfo  {journal} {Journal of Chemical Theory and Computation}\ }\textbf {\bibinfo {volume} {19}},\ \bibinfo {pages} {3429} (\bibinfo {year} {2023})}\BibitemShut {NoStop}%
\bibitem [{\citenamefont {Tian}\ \emph {et~al.}(2020)\citenamefont {Tian}, \citenamefont {Curnutte},\ and\ \citenamefont {Trcek}}]{tian2020rna}%
  \BibitemOpen
  \bibfield  {author} {\bibinfo {author} {\bibfnamefont {S.}~\bibnamefont {Tian}}, \bibinfo {author} {\bibfnamefont {H.~A.}\ \bibnamefont {Curnutte}},\ and\ \bibinfo {author} {\bibfnamefont {T.}~\bibnamefont {Trcek}},\ }\href@noop {} {\bibfield  {journal} {\bibinfo  {journal} {Molecules}\ }\textbf {\bibinfo {volume} {25}},\ \bibinfo {pages} {3130} (\bibinfo {year} {2020})}\BibitemShut {NoStop}%
\bibitem [{\citenamefont {Heidemann}\ and\ \citenamefont {Michelsen}(1995)}]{heidemann1995instability}%
  \BibitemOpen
  \bibfield  {author} {\bibinfo {author} {\bibfnamefont {R.~A.}\ \bibnamefont {Heidemann}}\ and\ \bibinfo {author} {\bibfnamefont {M.~L.}\ \bibnamefont {Michelsen}},\ }\href@noop {} {\bibfield  {journal} {\bibinfo  {journal} {Industrial \& Engineering Chemistry Research}\ }\textbf {\bibinfo {volume} {34}},\ \bibinfo {pages} {958} (\bibinfo {year} {1995})}\BibitemShut {NoStop}%
\bibitem [{\citenamefont {van Leuken}\ \emph {et~al.}(2023{\natexlab{b}})\citenamefont {van Leuken}, \citenamefont {van Benthem}, \citenamefont {Tuinier},\ and\ \citenamefont {Vis}}]{van2023predicting}%
  \BibitemOpen
  \bibfield  {author} {\bibinfo {author} {\bibfnamefont {S.~H.}\ \bibnamefont {van Leuken}}, \bibinfo {author} {\bibfnamefont {R.~A.}\ \bibnamefont {van Benthem}}, \bibinfo {author} {\bibfnamefont {R.}~\bibnamefont {Tuinier}},\ and\ \bibinfo {author} {\bibfnamefont {M.}~\bibnamefont {Vis}},\ }\href@noop {} {\bibfield  {journal} {\bibinfo  {journal} {Macromolecular Theory and Simulations}\ }\textbf {\bibinfo {volume} {32}},\ \bibinfo {pages} {2300001} (\bibinfo {year} {2023}{\natexlab{b}})}\BibitemShut {NoStop}%
\bibitem [{\citenamefont {Sollich}(2001)}]{sollich2001predicting}%
  \BibitemOpen
  \bibfield  {author} {\bibinfo {author} {\bibfnamefont {P.}~\bibnamefont {Sollich}},\ }\href@noop {} {\bibfield  {journal} {\bibinfo  {journal} {Journal of Physics: Condensed Matter}\ }\textbf {\bibinfo {volume} {14}},\ \bibinfo {pages} {R79} (\bibinfo {year} {2001})}\BibitemShut {NoStop}%
\bibitem [{\citenamefont {Koningsveld}(1969)}]{koningsveld1969phase}%
  \BibitemOpen
  \bibfield  {author} {\bibinfo {author} {\bibfnamefont {R.}~\bibnamefont {Koningsveld}},\ }\href@noop {} {\bibfield  {journal} {\bibinfo  {journal} {Pure and Applied Chemistry}\ }\textbf {\bibinfo {volume} {20}},\ \bibinfo {pages} {271} (\bibinfo {year} {1969})}\BibitemShut {NoStop}%
\bibitem [{\citenamefont {Koningsveld}(1970)}]{koningsveld1970preparative}%
  \BibitemOpen
  \bibfield  {author} {\bibinfo {author} {\bibfnamefont {R.}~\bibnamefont {Koningsveld}},\ }in\ \href@noop {} {\emph {\bibinfo {booktitle} {Fortschritte der Hochpolymeren-Forschung}}}\ (\bibinfo  {publisher} {Springer Berlin Heidelberg},\ \bibinfo {address} {Berlin, Heidelberg},\ \bibinfo {year} {1970})\ pp.\ \bibinfo {pages} {1--69}\BibitemShut {NoStop}%
\bibitem [{\citenamefont {Koningsveld}\ \emph {et~al.}(2001)\citenamefont {Koningsveld}, \citenamefont {Stockmayer},\ and\ \citenamefont {Nies}}]{koningsveld2001polymer}%
  \BibitemOpen
  \bibfield  {author} {\bibinfo {author} {\bibfnamefont {R.}~\bibnamefont {Koningsveld}}, \bibinfo {author} {\bibfnamefont {W.~H.}\ \bibnamefont {Stockmayer}},\ and\ \bibinfo {author} {\bibfnamefont {E.}~\bibnamefont {Nies}},\ }\href@noop {} {\emph {\bibinfo {title} {Polymer phase diagrams: a textbook}}}\ (\bibinfo  {publisher} {Oxford University Press},\ \bibinfo {year} {2001})\BibitemShut {NoStop}%
\bibitem [{\citenamefont {{\v{S}}olc}(1970)}]{solec1970cloud}%
  \BibitemOpen
  \bibfield  {author} {\bibinfo {author} {\bibfnamefont {K.}~\bibnamefont {{\v{S}}olc}},\ }\href@noop {} {\bibfield  {journal} {\bibinfo  {journal} {Macromolecules}\ }\textbf {\bibinfo {volume} {3}},\ \bibinfo {pages} {665} (\bibinfo {year} {1970})}\BibitemShut {NoStop}%
\bibitem [{\citenamefont {{\v{S}}olc}(1975)}]{vsolc1975cloud}%
  \BibitemOpen
  \bibfield  {author} {\bibinfo {author} {\bibfnamefont {K.}~\bibnamefont {{\v{S}}olc}},\ }\href@noop {} {\bibfield  {journal} {\bibinfo  {journal} {Macromolecules}\ }\textbf {\bibinfo {volume} {8}},\ \bibinfo {pages} {819} (\bibinfo {year} {1975})}\BibitemShut {NoStop}%
\bibitem [{\citenamefont {Tompa}(1949)}]{tompa1949phase}%
  \BibitemOpen
  \bibfield  {author} {\bibinfo {author} {\bibfnamefont {H.}~\bibnamefont {Tompa}},\ }\href@noop {} {\bibfield  {journal} {\bibinfo  {journal} {Transactions of the Faraday Society}\ }\textbf {\bibinfo {volume} {45}},\ \bibinfo {pages} {1142} (\bibinfo {year} {1949})}\BibitemShut {NoStop}%
\bibitem [{\citenamefont {Koningsveld}\ and\ \citenamefont {Staverman}(1967)}]{koningsveld1967liquid}%
  \BibitemOpen
  \bibfield  {author} {\bibinfo {author} {\bibfnamefont {R.}~\bibnamefont {Koningsveld}}\ and\ \bibinfo {author} {\bibfnamefont {A.}~\bibnamefont {Staverman}},\ }\href@noop {} {\bibfield  {journal} {\bibinfo  {journal} {Kolloid-Zeitschrift und Zeitschrift f{\"u}r Polymere}\ }\textbf {\bibinfo {volume} {220}},\ \bibinfo {pages} {31} (\bibinfo {year} {1967})}\BibitemShut {NoStop}%
\bibitem [{\citenamefont {Sollich}\ and\ \citenamefont {Cates}(1998)}]{sollich1998projected}%
  \BibitemOpen
  \bibfield  {author} {\bibinfo {author} {\bibfnamefont {P.}~\bibnamefont {Sollich}}\ and\ \bibinfo {author} {\bibfnamefont {M.~E.}\ \bibnamefont {Cates}},\ }\href@noop {} {\bibfield  {journal} {\bibinfo  {journal} {Physical Review Letters}\ }\textbf {\bibinfo {volume} {80}},\ \bibinfo {pages} {1365} (\bibinfo {year} {1998})}\BibitemShut {NoStop}%
\bibitem [{\citenamefont {Fasolo}\ and\ \citenamefont {Sollich}(2003)}]{fasolo2003equilibrium}%
  \BibitemOpen
  \bibfield  {author} {\bibinfo {author} {\bibfnamefont {M.}~\bibnamefont {Fasolo}}\ and\ \bibinfo {author} {\bibfnamefont {P.}~\bibnamefont {Sollich}},\ }\href@noop {} {\bibfield  {journal} {\bibinfo  {journal} {Physical Review Letters}\ }\textbf {\bibinfo {volume} {91}},\ \bibinfo {pages} {068301} (\bibinfo {year} {2003})}\BibitemShut {NoStop}%
\bibitem [{\citenamefont {Fasolo}\ and\ \citenamefont {Sollich}(2004)}]{fasolo2004fractionation}%
  \BibitemOpen
  \bibfield  {author} {\bibinfo {author} {\bibfnamefont {M.}~\bibnamefont {Fasolo}}\ and\ \bibinfo {author} {\bibfnamefont {P.}~\bibnamefont {Sollich}},\ }\href@noop {} {\bibfield  {journal} {\bibinfo  {journal} {Physical Review E—Statistical, Nonlinear, and Soft Matter Physics}\ }\textbf {\bibinfo {volume} {70}},\ \bibinfo {pages} {041410} (\bibinfo {year} {2004})}\BibitemShut {NoStop}%
\bibitem [{\citenamefont {Speranza}\ and\ \citenamefont {Sollich}(2003)}]{speranza2003isotropic2}%
  \BibitemOpen
  \bibfield  {author} {\bibinfo {author} {\bibfnamefont {A.}~\bibnamefont {Speranza}}\ and\ \bibinfo {author} {\bibfnamefont {P.}~\bibnamefont {Sollich}},\ }\href@noop {} {\bibfield  {journal} {\bibinfo  {journal} {The Journal of Chemical Physics}\ }\textbf {\bibinfo {volume} {118}},\ \bibinfo {pages} {5213} (\bibinfo {year} {2003})}\BibitemShut {NoStop}%
\bibitem [{\citenamefont {Wilding}\ \emph {et~al.}(2005)\citenamefont {Wilding}, \citenamefont {Sollich},\ and\ \citenamefont {Fasolo}}]{wilding2005finite}%
  \BibitemOpen
  \bibfield  {author} {\bibinfo {author} {\bibfnamefont {N.~B.}\ \bibnamefont {Wilding}}, \bibinfo {author} {\bibfnamefont {P.}~\bibnamefont {Sollich}},\ and\ \bibinfo {author} {\bibfnamefont {M.}~\bibnamefont {Fasolo}},\ }\href@noop {} {\bibfield  {journal} {\bibinfo  {journal} {Physical Review Letters}\ }\textbf {\bibinfo {volume} {95}},\ \bibinfo {pages} {155701} (\bibinfo {year} {2005})}\BibitemShut {NoStop}%
\bibitem [{\citenamefont {Patyukova}\ \emph {et~al.}(2021)\citenamefont {Patyukova}, \citenamefont {Xi},\ and\ \citenamefont {Wilson}}]{patyukova2021phase}%
  \BibitemOpen
  \bibfield  {author} {\bibinfo {author} {\bibfnamefont {E.}~\bibnamefont {Patyukova}}, \bibinfo {author} {\bibfnamefont {E.}~\bibnamefont {Xi}},\ and\ \bibinfo {author} {\bibfnamefont {M.~R.}\ \bibnamefont {Wilson}},\ }\href@noop {} {\bibfield  {journal} {\bibinfo  {journal} {Macromolecules}\ }\textbf {\bibinfo {volume} {54}},\ \bibinfo {pages} {2763} (\bibinfo {year} {2021})}\BibitemShut {NoStop}%
\bibitem [{\citenamefont {de~Souza}\ and\ \citenamefont {Stone}(2024)}]{deSouza2024exact}%
  \BibitemOpen
  \bibfield  {author} {\bibinfo {author} {\bibfnamefont {J.~P.}\ \bibnamefont {de~Souza}}\ and\ \bibinfo {author} {\bibfnamefont {H.~A.}\ \bibnamefont {Stone}},\ }\href@noop {} {\bibfield  {journal} {\bibinfo  {journal} {The Journal of Chemical Physics}\ }\textbf {\bibinfo {volume} {161}},\ \bibinfo {pages} {044902} (\bibinfo {year} {2024})}\BibitemShut {NoStop}%
\bibitem [{\citenamefont {Huggins}\ and\ \citenamefont {Okamoto}(1967)}]{huggins1967theoretical}%
  \BibitemOpen
  \bibfield  {author} {\bibinfo {author} {\bibfnamefont {M.~L.}\ \bibnamefont {Huggins}}\ and\ \bibinfo {author} {\bibfnamefont {H.}~\bibnamefont {Okamoto}},\ }in\ \href@noop {} {\emph {\bibinfo {booktitle} {Polymer Fractionation}}}\ (\bibinfo  {publisher} {Elsevier},\ \bibinfo {year} {1967})\ pp.\ \bibinfo {pages} {1--42}\BibitemShut {NoStop}%
\bibitem [{\citenamefont {Koningsveld}\ and\ \citenamefont {Staverman}(1968)}]{koningsveld1968liquid_calculation}%
  \BibitemOpen
  \bibfield  {author} {\bibinfo {author} {\bibfnamefont {R.}~\bibnamefont {Koningsveld}}\ and\ \bibinfo {author} {\bibfnamefont {A.}~\bibnamefont {Staverman}},\ }\href@noop {} {\bibfield  {journal} {\bibinfo  {journal} {Journal of Polymer Science Part A-2: Polymer Physics}\ }\textbf {\bibinfo {volume} {6}},\ \bibinfo {pages} {305} (\bibinfo {year} {1968})}\BibitemShut {NoStop}%
\bibitem [{\citenamefont {de~Souza}\ \emph {et~al.}(2025)\citenamefont {de~Souza}, \citenamefont {Jacobs},\ and\ \citenamefont {Stone}}]{GitCode}%
  \BibitemOpen
  \bibfield  {author} {\bibinfo {author} {\bibfnamefont {J.~P.}\ \bibnamefont {de~Souza}}, \bibinfo {author} {\bibfnamefont {W.~M.}\ \bibnamefont {Jacobs}},\ and\ \bibinfo {author} {\bibfnamefont {H.~A.}\ \bibnamefont {Stone}},\ }\href@noop {} {\bibinfo {title} {{Flory-Huggins Binodal Polydisperse}}},\ \bibinfo {howpublished} {\url{https://github.com/jpldesouza/FH-Binodal_polydisperse}} (\bibinfo {year} {2025})\BibitemShut {NoStop}%
\bibitem [{Not()}]{Note1}%
  \BibitemOpen
  \href@noop {} {}\bibinfo {howpublished} {Note that distributions of this type may be reasonably approximated by a normal distribution when $\lambda$ is large}\BibitemShut {NoStop}%
\bibitem [{\citenamefont {Chen}\ and\ \citenamefont {Jacobs}(2023)}]{chen2023emergence}%
  \BibitemOpen
  \bibfield  {author} {\bibinfo {author} {\bibfnamefont {F.}~\bibnamefont {Chen}}\ and\ \bibinfo {author} {\bibfnamefont {W.~M.}\ \bibnamefont {Jacobs}},\ }\href@noop {} {\bibfield  {journal} {\bibinfo  {journal} {Journal of Chemical Theory and Computation}\ } (\bibinfo {year} {2023})}\BibitemShut {NoStop}%
\bibitem [{{\relax DLMF}()}]{NIST:DLMF:1.10.vii}%
  \BibitemOpen
  {\relax DLMF},\ \href {https://dlmf.nist.gov/1.10.vii} {\bibinfo {title} {{\it NIST Digital Library of Mathematical Functions}, s 1.10(vii), {Lagrange Inversion Theorem}}},\ \bibinfo {howpublished} {\url{https://dlmf.nist.gov/1.10.vii}, Release 1.2.6 of 2026-03-15},\ \bibinfo {note} {f.~W.~J. Olver, A.~B. {Olde Daalhuis}, D.~W. Lozier, B.~I. Schneider, R.~F. Boisvert, C.~W. Clark, B.~R. Miller, B.~V. Saunders, H.~S. Cohl, and M.~A. McClain, eds.}\BibitemShut {Stop}%
\end{thebibliography}%

\end{document}